\documentclass{article}%

\pdfoutput=1

\usepackage{geometry}
\usepackage{amsmath, amsthm, amssymb}
\usepackage{changepage}
\usepackage{fancyhdr}
\usepackage{graphicx}
\usepackage{nameref}
\usepackage{xcolor}
\usepackage{bbm}
\usepackage{multicol}
\usepackage{enumerate}

\usepackage{threeparttable}
\usepackage{multirow}

\usepackage{nameref,hyperref}
\usepackage{array}

\newcolumntype{+}{!{\vrule width 2pt}}

\newlength\savedwidth

\newcommand\thickhline{\noalign{\global\savedwidth\arrayrulewidth\global\arrayrulewidth 2pt}%
\hline
\noalign{\global\arrayrulewidth\savedwidth}}

\newcommand{\ddt}[1]{\frac{\mathrm{d}#1}{\mathrm{dt}}}

\newcommand{\um}{\mathrm{\mu m}}
\newcommand{\ug}{\mathrm{\mu g}}

\newcommand{\Diskus}{Diskus\textsuperscript{\textregistered}}
\newcommand{\Turbohaler}{Turbohaler\textsuperscript{\textregistered}}

\newcolumntype{L}[1]{>{\raggedright\let\newline\\\arraybackslash\hspace{0pt}}m{#1}}
\newcolumntype{C}[1]{>{\centering\let\newline\\\arraybackslash\hspace{0pt}}m{#1}}
\newcolumntype{R}[1]{>{\raggedleft\let\newline\\\arraybackslash\hspace{0pt}}m{#1}}

\author{Niklas Hartung\textsuperscript{1*}, Jens Borghardt\textsuperscript{2}} \date{}
\title{A mechanistic framework for \emph{a priori} pharmacokinetic predictions of orally inhaled drugs}

\begin{document}

\maketitle

\begin{flushleft}
\textbf{1} Institute of Mathematics, University of Potsdam, Potsdam, Germany
\\
\textbf{2} Drug Discovery Sciences, Research Pharmacokinetics, Boehringer Ingelheim Pharma GmbH \& Co. KG, Biberach, Germany
\\
\bigskip

* niklas.hartung@uni-potsdam.de

\end{flushleft}

\begin{abstract}
The fate of orally inhaled drugs is determined by pulmonary pharmacokinetic processes such as particle deposition, pulmonary drug dissolution, and mucociliary clearance. Even though each single process has been systematically investigated, a quantitative understanding on the interaction of processes remains limited and therefore identifying optimal drug and formulation characteristics for orally inhaled drugs is still challenging. To investigate this complex interplay, the pulmonary processes can be integrated into mathematical models. However, existing modeling attempts considerably simplify these processes or are not systematically evaluated against (clinical) data. 
In this work, we developed a mathematical framework based on physiologically-structured population equations to integrate all relevant pulmonary processes mechanistically. A tailored numerical resolution strategy was chosen and the mechanistic model was evaluated systematically against data from different clinical studies. Without adapting the mechanistic model or estimating kinetic parameters based on individual study data, the developed model was able to predict simultaneously (i) lung retention profiles of inhaled insoluble particles, (ii) particle size-dependent pharmacokinetics of inhaled monodisperse particles, (iii) pharmacokinetic differences between inhaled fluticasone propionate and budesonide, as well as (iv) pharmacokinetic differences between healthy volunteers and asthmatic patients. 
Finally, to identify the most impactful optimization criteria for orally inhaled drugs, the developed mechanistic model was applied to investigate the impact of input parameters on both the pulmonary and systemic exposure. Interestingly, the solubility of the inhaled drug did not have any relevant impact on the local and systemic pharmacokinetics. Instead, the pulmonary dissolution rate, the particle size, the tissue affinity, and the systemic clearance were the most impactful potential optimization parameters. In the future, the developed prediction framework should be considered a powerful tool for identifying optimal drug and formulation characteristics.
\end{abstract}
\section*{Introduction}

Oral drug inhalation can result in high pulmonary drug exposure while maintaining low systemic exposure. Compared to other routes of administration, this can provide higher local pulmonary efficacy, while simultaneously reducing systemic adverse effects (``lung selectivity'') \cite{lipworth1999,borghardt2018,jennings1991}. Therefore, orally inhaled drugs are considered first-line therapy (amongst other treatment options) to treat respiratory diseases such as asthma bronchial or chronic obstructive pulmonary disease \cite{gold2015,gina2016}.

While qualitatively, the pharmacodynamic (PD) selectivity for the lung was previously investigated, a sound quantitative understanding about the pulmonary pharmacokinetics (PK) is still lacking. Specific pulmonary PK processes after oral drug inhalation were studied in detail, such as the pulmonary particle deposition \cite{longest2012,heyder1986,clark2012} or mucociliary clearance \cite{hofmann2003,hofmann2004}. For example, it is well understood that the central airway deposition increases with an increasing aerodynamic particle size \cite{heyder1986} and that the mucociliary clearance depends on the localization in the airways \cite{hofmann2004}. Hence, the impact of mucociliary clearance strongly depends on particle deposition patterns. However, in contrast to investigations related to the individual processes, the interplay of the many pulmonary PK processes has received less attention. A comprehensive quantitative understanding of how these processes contribute to pulmonary and systemic PK, and therefore to lung selectivity after drug inhalation, is often still lacking \cite{patton2007,borghardt2015,ruge2013,labiris2003}. Thus, identifying drug and formulation characteristics for orally inhaled drugs that maximize lung selectivity as well as long-lasting pulmonary efficacy is still challenging.

To gain a better understanding on the interplay of pulmonary PK processes, mechanistic modeling approaches can be applied. However, previous modeling approaches either reduced the given complexity or lack adequate model evaluation. For example, the mucociliary clearance was described as a first-order process \cite{weber2013,raut2019}. Other published population PK models did not differentiate between undissolved and dissolved drug and consider pulmonary drug absorption as a ``one-way process'', i.e. back flow of drug to the lungs from the systemic disposition is not considered \cite{borghardt2016,bartels2013,melin2017}. One mechanistic partial differential equation (PDE)-based model is available, which included all relevant pulmonary PK processes \cite{boger2018}. This model, however, was not evaluated against clinical data. Hence, to our knowledge no fully mechanistic model, with an adequate model evaluation based on clinical and in vitro data, is available. Consequently, there is currently no adequate framework to quantitatively identify the most impactful drug and formulation characteristics to achieve good lung selectivity.

\begin{figure}[!h]
\centering
\includegraphics[width=.9\textwidth]{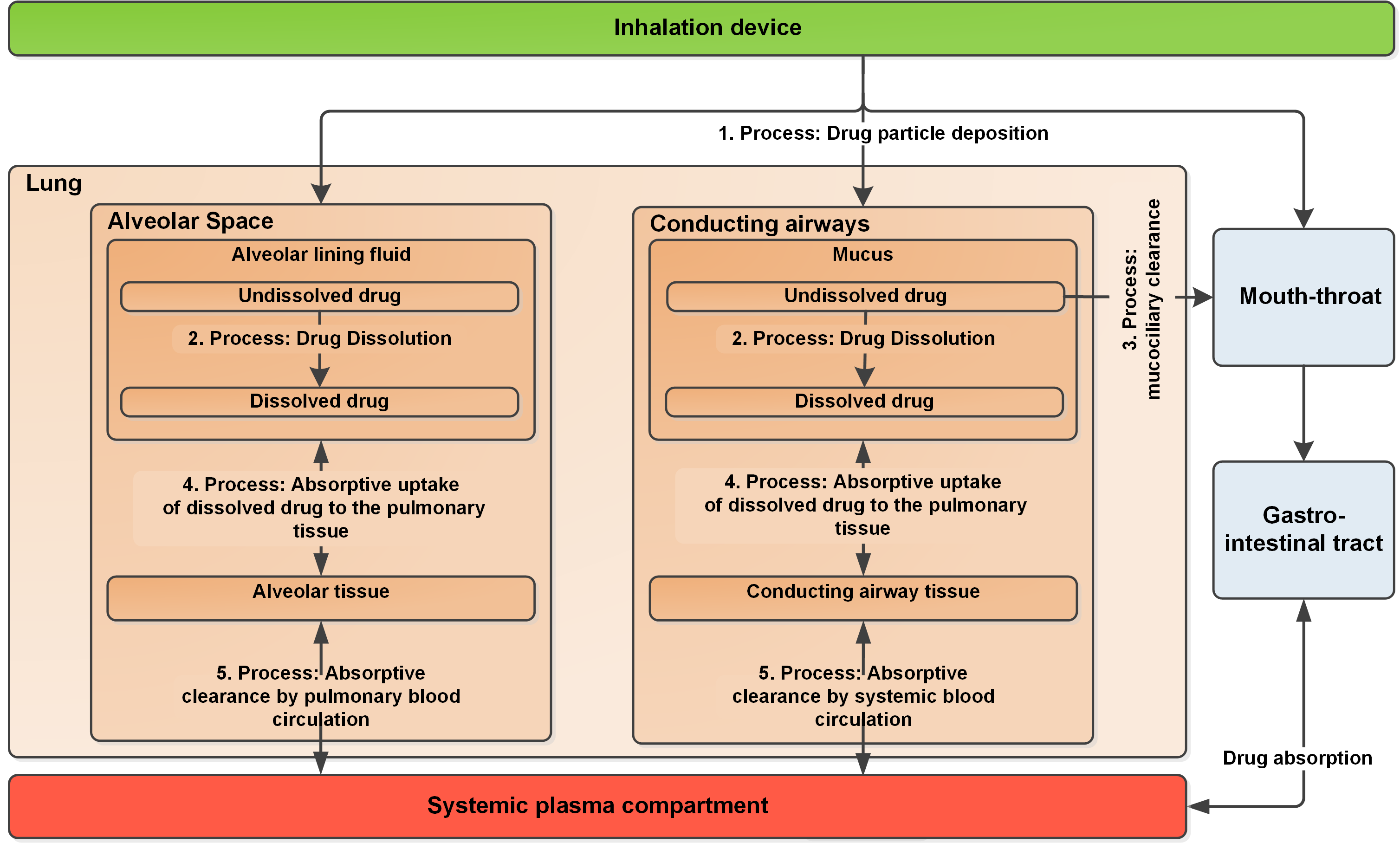}
\caption{\label{fig1} {\bf Overview of relevant pulmonary pharmacokinetic processes for orally inhaled drugs.}
Adapted and modified from \cite{borghardt2018}.}
\end{figure}

In this work, we aimed at developing such a mechanistic pulmonary PK model to capture the complexity of all relevant pulmonary PK processes (compare Fig~\ref{fig1}) and to determine which parameters are the most suitable optimization criteria to achieve optimal lung selectivity. The biggest mathematical challenge related to such a model is to adequately describe the joint effect of location-dependent nonlinear mucociliary clearance and particle size-dependent dissolution. To achieve this, a size- and location-structured PDE model was developed. The resulting PDE model was extensively evaluated, in particular based on clinical PK data for both budesonide and fluticasone propionate, as these inhaled drugs represent the clinically most studied compounds. Finally, a sensitivity analysis was performed to determine the most impactful drug and formulation characteristics and therefore potential optimization parameters to achieve a high lung selectivity. 

\section*{Models}

The mathematical model is introduced in a stepwise manner. First, the (sub)models describing the considered pulmonary PK processes are given. Next, the full PDE model it presented. The model parametrization is described in the Results section. Full details concerning derivations, numerical resolution, and additional model evaluations are given in \nameref{S1_Appendix}, as referenced below.

\subsection*{Modeling of pharmacokinetic processes in the lung}

\subsubsection*{Pulmonary particle deposition}

Since orally inhaled drugs are deposited in the lungs within a single breath, pulmonary drug deposition was considered as an instantaneous rather than a time-dependent process. Pulmonary particle deposition was simulated with the MPPD software \cite{mppd} according to the study design of each investigated study (i.e., for monodisperse particle size formulations as well as the specific particle size distributions of the \Diskus and \Turbohaler devices, respectively \cite{tamura2012}). To simulate deposition patterns for asthmatic patients, who are characterized by a more central deposition compared to healthy volunteers \cite{kim1997,darquenne2012}, we corrected the deposition patterns in healthy volunteers based on scintigraphy data reported in \cite{usmani2005}. A full account of input parameters to predict the deposition patterns and the adaption for asthmatic patients is provided in \nameref{S1_Appendix}(Section 4). 

These predictions generated aerodynamic particle size- and lung generation-resolved deposition patterns. The aerodynamic particle size (the size of a water particle experiencing the same aerodynamic forces as the considered particle) determines the deposition characteristics of the inhaled particles \cite{smith2008}. In contrast, the real (geometric) size of an inhaled particle is relevant for dissolution processes \cite{noyes1897}. To convert aerodynamic to geometric particle sizes, which is more relevant for dissolution characteristics, we assumed a spherical shape of particles and considered the relationship

\[d_\text{geom} = d_\text{aero} \sqrt{\frac{\rho_\text{water}}{\rho_\text{substance}}},\]
where $d_\text{aero}$ and $d_\text{geom}$ are aerodynamic and geometric particle diameters, respectively; $\rho_\text{water}$ and $\rho_\text{substance}$ denote density of water and the considered inhaled substance, respectively \cite{hinds1999}.

In a post-processing step, the (geometric) particle size- and lung generation-resolved deposition patterns were projected onto the computational grid, ensuring conservation of the number of molecules (full details are given in \nameref{S1_Appendix}, Section 2.5.1).

\subsubsection*{Mucociliary clearance}

The mucociliary clearance process was parameterized based on a model for  mucociliary clearance published by Hofmann and Sturm (see \nameref{S1_Appendix}, Section~1.2 for details) \cite{hofmann2004}. In agreement with clinical data, mucociliary clearance of undissolved particles only depends on the particle location, not on (geometric) particle size \cite{smith2008}:

\[\lambda_\text{mc}(x) = 0.8791 \frac{\text{cm}}{\text{min}}\cdot \left( \frac{r^\text{br}(x)}{1~\text{cm}}\right)^{2.808},\] 
where $r^\text{br}(x)$ represents the radius of the conducting airways at location $x$.

\subsubsection*{Pulmonary drug dissolution}

The dissolution of particles in the pulmonary lining fluids was based on an adapted version of the Noyes-Whitney equation \cite{noyes1897}: 

\[d(s,C_\text{flu}) = \frac{4\pi \,k_\text{diss}}{(\frac{4}{3}\pi)^{1/3}\,\rho}\cdot \left(1-\frac{C_\text{flu}}{C_s}\right) \cdot s^{1/3},\]
where $s$ denotes the particle volume, $\rho$ the particle density, $C_s$ the saturation solubility, $k_\text{diss}=D\cdot C_s$ the maximum dissolution rate ($D$= diffusivity), and $C_\text{flu}$ the local concentration of dissolved drug in the lining fluid. A derivation of this equation from the Noyes-Whitney equation, assuming spherical particle geometry, is provided in \nameref{S1_Appendix}, Section~1.1.
To represent the difference in fluid composition between conducting airways and the alveolar space, in particular in terms of fluid viscosity, different dissolution rate constants ($k_\text{diss}^\text{br}$/$k_\text{diss}^\text{alv}$) were assumed in these two regions, leading to dissolution models $d^\text{br}$ and $d^\text{alv}$, respectively.

\subsubsection*{Absorption into the lung tissues}

After drug dissolution in the pulmonary lining fluids, the drug is absorbed through the airway epithelia into the lung tissue of the respective airway generation or the alveolar space. Based on reported negligible to tenfold lower albumin concentrations in epithelial ling fluids in the lung compared to plasma \cite{rennard1986,reifenrath1973,adams1963}, the absorption rate is calculated assuming no drug binding in the lung lining fluids: 

\[k_a = P_\text{app}\cdot \text{SA} \cdot \left(C_\text{flu} - \frac{C_\text{tis}}{K_\text{pu,tis}}\right),\]
where $k_a$ denotes the absorption rate, $P_\text{app}$ the effective permeability, SA the airway surface area, and $K_\text{pu,tis}$ describes the lung-to-unbound plasma partition coefficient.	

\subsubsection*{Systemic disposition}

The systemic disposition models for both budesonide and fluticasone propionate were based on available literature information after intravenous administration and oral administration (to include the oral bioavailability of swallowed drug). In contrast to many less mechanistic PK models, the backflow of drug from the systemic circulation into the lung was mechanistically included in the PDE-based PK model. 

\subsection*{Partial differential equation model for orally inhaled drugs}

\subsubsection*{Model equations}

To mechanistically combine the considered pulmonary processes in the lung (lung deposition, mucociliary clearance, pulmonary dissolution, pulmonary absorption to the lung tissue and distribution between lung tissue and plasma), we adopted the framework of physiologically structured population models (PSPMs)  \cite{metz1986}. In this class of PDE models, the time evolution of a density is described over a state space through a set of processes that modify the state. 

To describe the fate of undissolved particles deposited in the lung, we considered (i) a PSPM with size and location structure in the conducting airways and (ii) a PSPM with size structure in the alveolar space. In these models, size $s$ represents the geometric volume of particles, and location $x$ (length unit) the position along all conducting airways, between trachea and terminal bronchioles. The state $(x,s)$ of a particle is changed by  mucociliary clearance (impacts on $x$) and pulmonary dissolution (impacts on $s$).

\begin{figure}[!h]
\centering
\includegraphics[width=.9\textwidth]{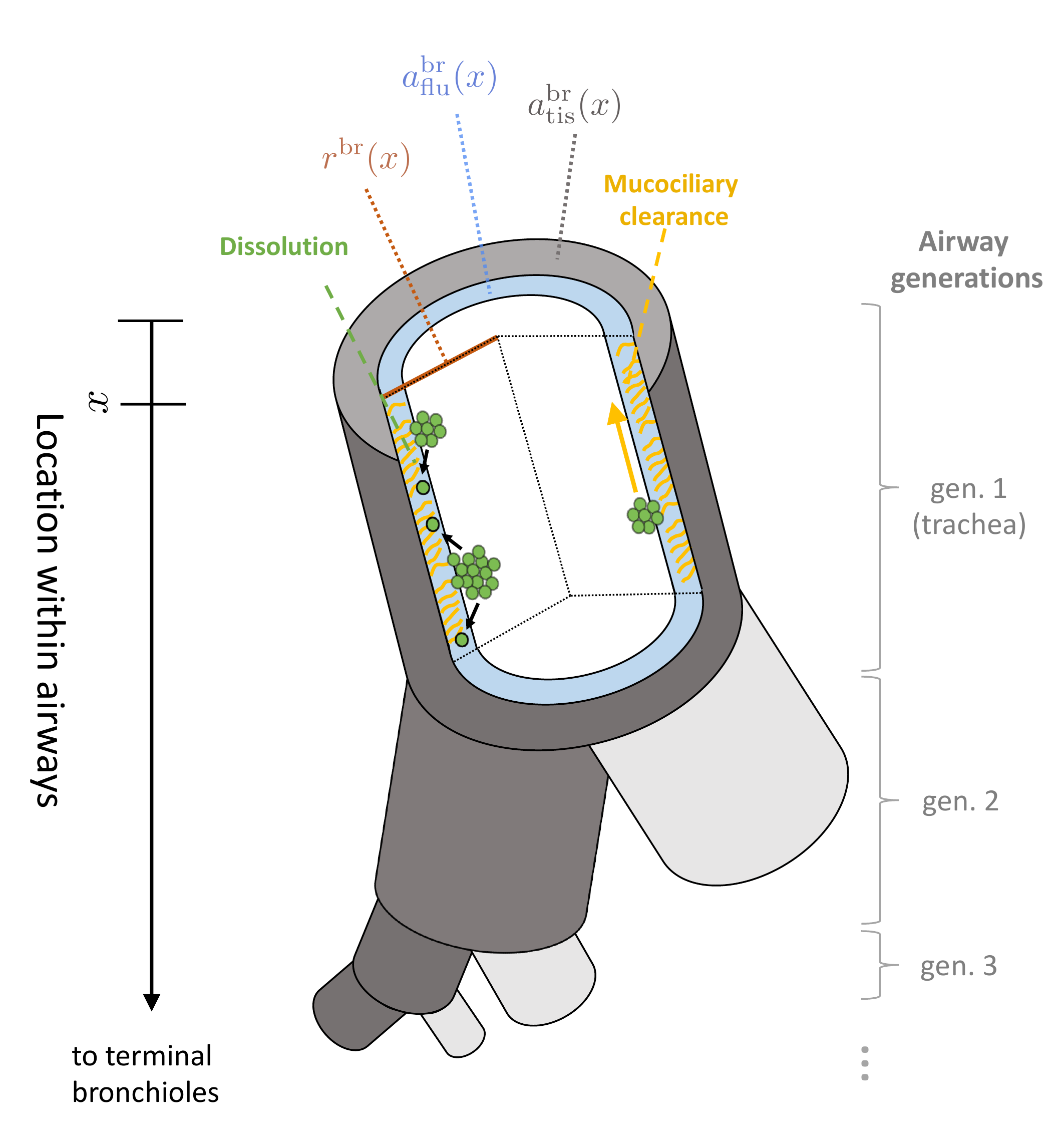}
\caption{{\bf Geometry of and key processes in conducting airways represented in the mathematical model.}
A representative airway (dark grey) is considered in the mathematical model. At each location $x$ (distance from throat) within this airway, we consider a cylindrical lung model, consisting of concentric layers of airway (with radius $r^\text{br}(x)$), lung lining fluid (with cross-sectional area $a_\text{flu}^\text{br}(x)$), and lung tissue (with cross-sectional area $a_\text{tis}^\text{br}(x)$), respectively. Each drug particle (green) is characterized by its location and size. Over time, particles are moved upwards by mucociliary clearance (yellow) and dissolve into the airway lining fluid (black arrows).
}
\label{process_illustrations}
\end{figure}

The PSPMs were coupled to differential equations describing the PK of dissolved drug molecules in lung lining fluids and lung tissues (similar to \cite{boger2016}) and published systemic disposition kinetics \cite{weber2013}. The full set of equations is stated below, a simplified outline of the underlying geometry is provided in Fig~\ref{process_illustrations} and a a detailed derivation of the full model from the separate PK processes is given in \nameref{S1_Appendix}~(Sections 1.3-1.5).

For ease of legibility, the following abbreviations are used as sub-/superscripts in the equations: br (bronchial, i.e., conducting airways), alv (alveolar space), sys (systemic), sol (solid, i.e., undissolved), flu (fluid, i.e., dissolved), tis (lung tissue), ctr (central), per (peripheral), mc (mucociliary clearance).

The size- and location-structured PSPM for the density of inhaled particles in suspension in the conducting airways reads
\begin{align*}
&\partial_t\rho^\text{br}(t,x,s) + \partial_x\big[\lambda_\text{mc}(x)\rho^\text{br}(t,x,s)\big] - \partial_s\big[d^\text{br}(s,C_\text{flu}^\text{br}(t,x))\rho^\text{br}(t,x,s)\big] = 0\\
& \rho^\text{br}(0,x,s) = \rho_0^\text{br}(x,s) \qquad\qquad \text{(see \nameref{S1_Appendix}, Sections~2.5.1 and~4)}\end{align*}
assuming zero inflow boundary conditions (i.e., no additional source of drug in the conducting airways after dosing).
This PDE was complemented by differential equations for the concentration of dissolved drug in lining fluids and lung tissue at a particular airway location $x$:
\begin{align*}
a_\text{flu}^\text{br}(x)\partial_t C_\text{flu}^\text{br}(t,x) &= \underbrace{\int\limits_0^{s_\text{max}}d^\text{br}(s,C_\text{flu}^\text{br}(t,x))\rho^\text{br}(t,x,s)\text{ds}}_{\text{dissolution}} - \underbrace{2\pi r^\text{br}(x)P_\text{app} \, \left(C_\text{flu}^\text{br}(t,x) - \frac{C_\text{tis}^\text{br}(t,x)}{K_\text{pu,tis}}\right)}_\text{tissue uptake}\\
a_\text{tis}^\text{br}(x)\partial_t C_\text{tis}^\text{br}(t,x) &=  2\pi r^\text{br}(x)P_\text{app} \, \left(C_\text{flu}^\text{br}(t,x) - \frac{C_\text{tis}^\text{br}(t,x)}{K_\text{pu,tis}}\right) - \underbrace{q^\text{br}(x)\left(\frac{\text{BP}\,C_\text{tis}^\text{br}(t,x)}{K_\text{p,tis}} - C_\text{ctr}^\text{sys}(t)\right)}_\text{systemic uptake}
\end{align*}
both with zero initial conditions.

In the alveolar space, the size-structured PSPM for the density of inhaled particles in suspension reads
\begin{align*}
&\partial_t\rho^\text{alv}(t,s) - \partial_s\big[d^\text{alv}(s,C_\text{flu}^\text{alv}(t))\rho^\text{alv}(t,s)\big] = 0\\
& \rho^\text{alv}(0,s) = \rho_0^\text{alv}(s) \qquad\qquad \text{(see \nameref{S1_Appendix}, Sections~2.5.1 and~4)}
\end{align*}
Again, zero inflow boundary conditions were assumed (no additional source of drug in the alveolar space after dosing) the  and the PDE is complemented by differential equations for the concentration of dissolved drug in alveolar lining fluids and alveolar lung tissue:
\begin{align*}
V_\text{flu}^\text{alv} \ddt{C_\text{flu}^\text{alv}}(t) &= \int\limits_0^{s_\text{max}}d^\text{alv}(s,C_\text{flu}^\text{alv}(t))\rho^\text{alv}(t,s)\text{d}s - P_\text{app} \,\text{SA}^\text{alv} \left(C_\text{flu}^\text{alv}(t) - \frac{C_\text{tis}^\text{alv}(t)}{K_\text{pu,tis}}\right)\\
V_\text{tis}^\text{alv}\ddt{C_\text{tis}^\text{alv}}(t) &=  P_\text{app} \, \text{SA}^\text{alv} \left(C_\text{flu}^\text{alv}(t) - \frac{C_\text{tis}^\text{alv}(t)}{K_\text{pu,tis}}\right) - Q^\text{alv}\left(\frac{\text{BP}\,C_\text{tis}^\text{alv}(t)}{K_\text{p,tis}} - C^\text{sys}_\text{ctr}(t)\right)
\end{align*}
with zero initial conditions.

The equations describing the PK in the conducting airways and the alveolar space are coupled through the systemic circulation:
\begin{align*}
V_\text{ctr} \ddt{C_\text{ctr}^\text{sys}}(t) &= \underbrace{\int\limits_0^{x_\text{TB}}q^\text{br}(x) \left(\frac{\text{BP}\,C_\text{tis}^\text{br}(t,x)}{K_\text{p,tis}} - C_\text{ctr}^\text{sys}(t)\right)\text{d}x}_{\text{exchange with conducting airways}} +
 \underbrace{Q^\text{alv} \left(\frac{\text{BP}\,C_\text{tis}^\text{alv}(t)}{K_\text{p,tis}} - C_\text{ctr}^\text{sys}(t)\right)}_{\text{exchange with alveolar space}}\\
&\quad - Q^\text{sys} \big(C_\text{ctr}^\text{sys}(t) - C_\text{per}^\text{sys}(t)\big) - \text{CL}\cdot C_\text{ctr}^\text{sys}(t)\\
V_\text{per}\ddt{C_\text{per}^\text{sys}}(t) &=  Q^\text{sys} \big(C_\text{ctr}^\text{sys}(t) - C_\text{per}^\text{sys}(t)\big)
\end{align*}
The following expressions appear in these equations:
\begin{itemize}
\item $x$ is the location within a prototypical airway, varying from 0 (trachea, corresponding to airway generation 1) to  $x_\text{TB}$ (terminal bronchioles, corresponding to airway generation 16). 
\item $s$ is the geometric particle volume, varying between 0 and $s_\text{max}$ (device- and formulation-specific maximum particle size deposited)
\item $\rho^\text{br}$/$\rho^\text{alv}$ are the PSPM densities, with units $\frac{\text{number of particles}}{\text{mL}\,\cdot\, \text{cm}}$ and $\frac{\text{number of particles}}{\text{mL}}$, respectively
\item $C_\text{z}^\text{y}$ is the concentrations of dissolved drug in lining fluid ($\text{z}=\text{flu}$) or lung tissue ($\text{z}=\text{tis}$) in a particular location of the conducting airways ($\text{y}=\text{br}$) or in the alveolar space ($\text{y}=\text{alv}$)
\item $d^\text{br}$ / $d^\text{alv}$ are the dissolution rates in conducting airways / alveolar space, depending on particle size $s$ and concentration of already dissolved drug
\item $\lambda_\text{mc}$ is the mucociliary clearance in the conducting airways, assumed to depend only on location $x$, not on (geometric) particle size $s$.
\item $P_\text{app}$ is the apparent permeability of the drug
\item $\text{SA}^\text{alv}$ is the surface area of the alveolar space
\item $r^\text{br}(x)$ is the airway radius (including lining fluid) at location $x$ (see Fig~\ref{process_illustrations})
\item $K_\text{p,tis}$ / $K_\text{pu,tis}$ are the lung-to-plasma and  lung-to-unbound plasma partition coefficients, respectively
\item $\text{BP}$ is the blood-to-plasma ratio of the drug
\item $a_\text{flu}^\text{br}(x)$ / $a_\text{tis}^\text{br}(x)$ is the cross-sectional area of lung lining fluid / lung tissue at location $x$ within the conducting airways
\item $q^\text{br}(x)$ is the location-resolved blood flow (see section~\nameref{sec:parametrization} below)
\end{itemize}

\subsubsection*{Numerical resolution}
 
To solve the mathematical model numerically, we employed an upwind discretization of the PSPMs \cite{eymard2000} together with an implicit discretization of all linear processes (MCC, absorption, systemic processes) \cite{courant1967,butcher2003}. The fluxes across PDEs (mucociliary elevator and dissolved / absorbed drug) were discretized ensuring that all conservation laws were fulfilled at the discrete level. The discretized model and all analyses were implemented in MATLAB~R2018b \cite{matlab}. A full description of the discretization scheme is given in \nameref{S1_Appendix} (Section 2) and the MATLAB implementation is provided as \nameref{S1_File}.

\section*{Results}

\subsection*{Key findings from literature review}

As a first step, PK studies for both budesonide and fluticasone propionate were identified. These drugs were selected as they represent the most studied inhaled drugs for which the interplay between pulmonary deposition, pulmonary dissolution, mucociliary clearance, as well as pulmonary absorption has been systematically discussed \cite{borghardt2018,darquenne2012}. In total, ten different clinical PK studies on these drugs were identified (see \nameref{S1_Table}).   

After reviewing all PK studies, we identified two important aspects. First, the area under the curve reported by Usmani et al. \cite{usmani2014} could not be reproduced considering the systemic clearance for fluticasone propionate reported by Mackie et al. \cite{mackie1996}. Even in the most extreme and certainly unrealistic assumptions, namely with 100\% of inhaled drug particles deposited in the lungs and no mucociliary clearance, the systemic AUC would still be at least 25\% lower than reported (see calculation in \nameref{S1_Appendix}, Section 3.3).

Second, there is a considerable between-study variability in reported (dose-normalized) systemic drug exposure (same drug, comparable dose, comparable patient population, same inhalation device). For example, both M\"ollmann et al. \cite{moellmann2001} and Harrison and Tattersfield \cite{harrison2003} investigated the systemic PK after budesonide inhalation with the \Turbohaler for slightly different doses of 1000~$\ug$, and 1200~$\ug$. The reported dose-normalized C\textsubscript{max} and AUC\textsubscript{0-Inf} values varied by more than twofold. In contrast, the relative shape of the PK profiles, which are not dependent on the absolute plasma concentrations, were in good agreement between both studies. A full summary of exposure metrics is given in \nameref{S1_Table}.

Based on these findings, we decided that predicting the absolute plasma concentrations of one single selected PK study is not meaningful or could even result in a selection bias. Instead, PK studies with multiple study arms, which allow for a direct within-study comparison of different PK profiles, were considered (i.e., studies with only a single investigated drug, a single inhaled particle size and a single investigated population were not included). A short overview of the reviewed PK studies, including a comment on why specific studies were considered, can be found in \nameref{S1_Table}. In summary, the selected PK studies comprised the following aspects relevant for model building and model evaluation: (i) the lung retention profiles for insoluble particles \cite{smith2008}, (ii) the impact of different particle sizes on the systemic PK \cite{usmani2014}, (iii) different systemic PK profiles after inhalation of either budesonide or fluticasone propionate \cite{moellmann2001,harrison2003}, and (iv) different systemic PK profiles between healthy volunteers and asthmatic patients \cite{harrison2003}.

\subsection*{Model parametrization}\label{sec:parametrization}

The PDE-based model was not adapted to individual studies, i.e., no (pharmacokinetic) parameters were estimated based on the studies which were used for model evaluation. Instead, the pulmonary part of the PDE model was fully parametrized based on physiological and drug-specific \emph{in vitro} data (\emph{a priori} predictions). 

Both pulmonary drug deposition and mucociliary clearance were considered as drug-independent generic processes based on particle size and airway characteristics alone, not requiring any drug-specific parameters. Drug-specific parameters, such as the maximum dissolution rate ($k_\text{diss}$), as well as drug solubility in pulmonary lining fluids were either based on literature information or in-house data on \emph{in vitro} dissolution and solubility. No direct comparison between alveolar and mucus dissolution kinetics could be retrieved from literature or in-house data. Therefore, a 5-fold decrease of $k_\text{diss}$ in the conducting airways compared to the alveolar space was assumed for all model-based simulations. A comprehensive list of parameter values is given in Table~\ref{tab:physiology} (physiological parameters) and Table~\ref{tab:drugdata} (drug-specific parameters). 

\begin{table}[!ht]
\centering
\begin{threeparttable}[b]
\caption{
\label{tab:physiology}
{\bf Physiological parameters.}}
\begin{tabular}{|r+l|l|}
\hline
\bf Parameter & \bf Symbol(s)  & \bf Value \\ \thickhline
Perfusion of conducting airways & $Q^\text{br}$  & 7.8~L/h~\tnote{\#1} \\ \hline
Perfusion of alveolar space     & $Q^\text{alv}$ & 312~L/h~\tnote{\#2}\\ \hline
Bronchial tissue volume & $V^\text{br}_\text{tis}$  & 144~mL~\tnote{\#3} \\ \hline
Alveolar tissue volume  & $V^\text{alv}_\text{tis}$ & 388~mL~\tnote{\#3} \\ \hline
Alveolar fluid volume & $V^\text{alv}_\text{flu}$ & 36~mL \cite{fronius2012} \\ \hline
Alveolar surface area & $\text{SA}^\text{alv}_\text{flu}$ & $130 \text{~m}^2$ \cite{weibel2009} \\ \hline\hline
Location-resolved parameters & $r^\text{br}$, $a^\text{br}_\text{flu}$, $a^\text{br}_\text{tis}$,$q^\text{br}$ & see main text \\ \hline 
\end{tabular}
\begin{tablenotes}
\item [\#1] calculated based on 2.5~\% of cardiac output \cite{gaohua2015}
\item [\#2] equal to cardiac output, taken from \cite[Table~22]{brown1997}
\item [\#3] computed from lung tissue weight of 532~g \cite{brown1997}, assuming a tissue density of 1~g/mL and 27~\% central/73~\% peripheral lung tissue weight fraction as in \cite[Supplement]{boger2016}.
\end{tablenotes}
\end{threeparttable}

\begin{flushleft} Summary of physiological parameters obtained from literature.
\end{flushleft}
\end{table}

\begin{table}[!ht]
\begin{adjustwidth}{-2.25in}{0in} 
\centering
\begin{threeparttable}[b]
\caption{
\label{tab:drugdata}
{\bf Drug-specific parameters.}}
\begin{tabular}{|r+l|l|l|}
\hline
\bf Parameter & \bf Symbol & \bf Fluticasone propionate & \bf Budesonide~~~~~~~~~\\ \thickhline
Central volume of distribution & $V_\text{ctr}$  & 31~L \cite{weber2013} & 100~L \cite{weber2013}\\ \hline
Peripheral volume of distribution & $V_\text{per}$  & 613~L \cite{weber2013} & 153~L \cite{weber2013}\\ \hline
Clearance & CL & 73~L/h \cite{weber2013} & 85~L/h \cite{weber2013}\\ \hline
Intercompartmental clearance & $Q^\text{sys}$ & 55.2~L/h \cite{weber2013} & 1701~L/h \cite{weber2013}\\ \hline
Oral bioavailability (of swallowed drug) & $F_\text{oral}$  & 0 \% \cite{weber2013} & 11 \% \cite{weber2013} \\ \hline
Absorption rate constant from GI tract & $k_a$  & -- & 0.45~1/h \cite{weber2013}\\ \hline
Fraction unbound in plasma & $f_\text{u,plasma}$ & 1.16 \%\tnote{\#1} & 16.1 \%\tnote{\#1} \\ \hline
Lung:plasma partition coefficient & $K_\text{p,tis}$ & 2.47\tnote{\#2} & 8 \cite{vandenbosch1993} \\ \hline
Permeability & $P_\text{app}$ & $92.6\cdot 10^{-6}$~cm/s\tnote{\#1} & $5.33\cdot 10^{-6}$~cm/s\cite{borchard2002} \\ \hline
Blood:plasma ratio & $\text{BP}$ & 1.83\tnote{\#1} & 0.8 \cite{pulmicort2008,szefler1999}\\ \hline
Molecular weight & MW & 500.57 & 430.53\\ \hline
Density & $\rho$ & 1.43~mol/L & 3.02~mol/L\\ \hline
Solubility & $C_s$ & $12.0~\mathrm{\mu}\text{M}$\tnote{\#1} & $69.8~\mathrm{\mu}\text{M}$\tnote{\#1} \\ \hline 
Maximum dissolution rate & $k_\text{diss}^\text{alv}$ & $6.17\cdot 10^{-5}~\frac{\text{nmol}}{\text{cm} \cdot \text{min}}$~\tnote{\#3} & $3.3\cdot 10^{-4}~\frac{\text{nmol}}{\text{cm} \cdot \text{min}}$~\tnote{\#3} \\ \hline \hline
Inhalation device-specific parameters &  \multicolumn{3}{|c|}{see \nameref{S1_Appendix}, Section~4}\\ \hline
\end{tabular}
\begin{tablenotes}
\item [\#1] in-house data: fraction unbound was determined with an \emph{in vitro} binding assay as described in \cite{cui2019}, permeability was determined based on an \emph{in vitro} permeability assay with Calu cells, with assay conditions as described for MDCK II cells in \cite{cui2019}. The \emph{in vitro} assay setup for determining  Blood:Plasma ratio and drug solubility in surfactant-containing media is described in \nameref{S1_Appendix}, Section 5.
\item [\#2] calculated based on $f_\text{u,plasma}$ (in-house data) and rat lung slice binding \cite{backstrom2016b}
\item [\#3] determined from in vitro dissolution data from \cite{rohrschneider2015} (see \nameref{S1_Appendix}, Section~3.1 for full details).
\end{tablenotes}
\end{threeparttable}

\begin{flushleft} For both drugs, two-compartment systemic PK models proposed in the literature were used.
\end{flushleft}
\end{adjustwidth}
\end{table}

To achieve a location-resolved parametrization of the conducting airways, we used generation-specific anatomical data of the conducting airways from \cite{yeh1980}, namely the length $l(g)$ and radius $r(g)$ of each airway generation $g$. From these values, location-resolved blood flows and cross-sectional lining fluid and lung tissue areas were calculated by assuming the following:
\begin{itemize}

\item Using length of airway generations, we determined a continuous representation $r^\text{br}(x)$ of the airway radius by linear interpolation between airway centerpoints.

\item We assessed literature data on lining fluid height $h_\text{flu}^\text{br}(x)$ for different airway generations and found an appropriate linear location-to-height of lining fluid-relationship (see \nameref{S1_Appendix}, Section 3.2, for details). Using the cylindrical geometry assumption depicted in Fig~\ref{process_illustrations}B,
$a_\text{flu}^\text{br}(x)$ could be determined from $h_\text{flu}^\text{br}(x)$ and the airway radius $r^\text{br}(x)$ via 
\[a_\text{flu}^\text{br}(x) = \pi\Big( r^\text{br}(x)^2 - \big(r^\text{br}(x)-h_\text{flu}^\text{br}(x)\big)^2\Big).\]

\item We assumed the cross-sectional area of conducting airway tissue $a_\text{tis}^\text{br}(x)$ to be proportional to cross-sectional lining fluid area $a_\text{flu}^\text{br}(x)$, with proportionality constant determined by the known total tissue volume of the central lung $V^\text{br}_\text{tis}$, i.e., via the relation $\int_0^{x_\text{TB}} a_\text{tis}^\text{br}(x)\mathrm{d}x = V^\text{br}_\text{tis}$.

\item We assumed a homogeneous perfusion of drug tissue within the conducting airway tissue, i.e., 
a location-resolved blood flow $q^\text{br}(x)$ proportional to $a_\text{tis}^\text{br}(x)$ and matching the total blood flow in the central lung $Q^\text{br}$, i.e. such that $\int_0^{x_\text{TB}} q^\text{br}(x) \mathrm{d}x=Q^\text{br}$.

\end{itemize}
We emphasize that the pulmonary PDE model was fully parametrized based on in vitro and physiological data, not fitted to the clinical data described in section \nameref{evaluation} below. A single adaptation was done based on physiological reasoning since no quantitative literature information could be retrieved, namely a 5-fold decrease of dissolution rate in the conducting airways compared to the alveolar space. The reason for this adapted dissolution rate constants is that the epithelial lining fluid in the conducting airways --the mucus-- contains a lower concentration of surfactants (which facilitate dissolution), compared to the alveolar lining fluid. In addition, the upper layer of the mucus is characterized by a high viscosity, which can also lead to a slower dissolution in comparison to the alveolar space.

\subsection*{Model evaluation}\label{evaluation}

\begin{figure}[!h]
\centering
\includegraphics[width=\textwidth,trim = 20mm 40mm 20mm 30mm,clip]{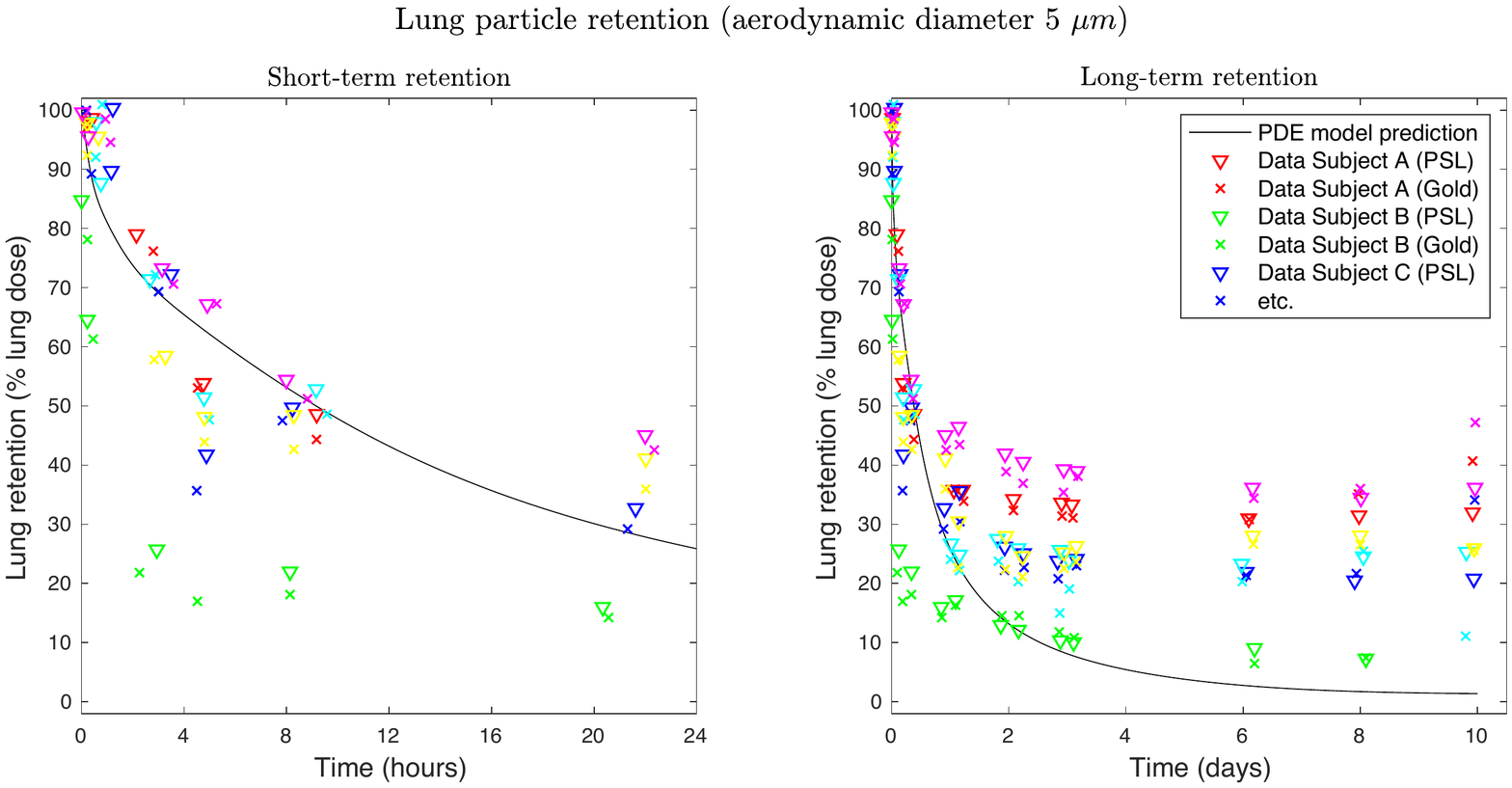}
\caption{{\bf Pulmonary retention profiles of inhaled insoluble particles.}
Pulmonary retention of inhaled monodisperse 5~$\mu$m-sized (aerodynamic diameter) gold and polystyrene (PSL) particles. The amount retained is described as a fraction of the initially deposited lung dose. Left: retention-time profile over 24~h, right: retention-time profile over 10~days. Data points: digitized data from six subjects from \cite{smith2008}; solid line: model-based predictions of lung retention.
}
\label{lung_retention}
\end{figure}

The mathematical model was evaluated in a stepwise approach. The first evaluation of the PDE-based inhalation PK model was based on a simulation of inhaled gold / polystyrene particles. As these particles do not dissolve in the pulmonary lining fluids, the interplay of deposition and mucociliary clearance can be evaluated independent of other pulmonary PK processes such as pulmonary dissolution or drug absorption. The initial particle retention was well described with $47\%$ of the deposited particles retained over 8~h and $26\%$ retained over 24~h (Fig~\ref{lung_retention}, left). However, the retention after 48~h was underpredicted, i.e. the data indicated a fraction of 7--36\% not being cleared from the lung, whereas the simulation indicated less than $5\%$ retention (Fig~\ref{lung_retention}, right). 

After evaluating the interplay of pulmonary deposition and mucociliary clearance, the pulmonary dissolution process was evaluated based on data from inhaled monodisperse drug formulations. In this evaluation step, the systemic PK of 1.5, 3, and 6 $\um$-sized particles (aerodynamic diameter) were simulated for the slowly dissolving inhaled drug fluticasone propionate. Simulation results were compared to the determined AUC\textsubscript{0-12h}, C\textsubscript{max}, and T\textsubscript{max} published by Usmani et al. As explained above, the absolute exposure metrics stated in the publication could not be reproduced. Rather than through goodness of prediction of \emph{absolute} exposure measures, we therefore evaluated the model by comparing the \emph{relative} change of exposure metrics across the three considered particle sizes. Of the model-predicted exposure metrics AUC\textsubscript{0-12h} and C\textsubscript{max}, 67\% were within 2-fold and 83\% within 3-fold of the reported ratios (compare Table~\ref{tab:usmani}). The predicted 1.5~$\um$~:~3~$\um$ T\textsubscript{max} ratio matched the experimental data well, however the other predicted T\textsubscript{max} ratios showed larger discrepancies due to a predicted very flat concentration-time profile for 6~$\um$ particles.

\begin{table}[!ht]
\centering
\begin{threeparttable}[b]
\caption{\label{tab:usmani}
{\bf Evaluation of model predictions for different particle sizes.}}
\begin{tabular}{|r+c|c|c|c|c|c|}
\hline
\bf Exposure  & \multicolumn{2}{|c|}{\bf AUC\textsubscript{0-12h}} & \multicolumn{2}{|c|}{\bf  C\textsubscript{max}}& \multicolumn{2}{|c|}{\bf  T\textsubscript{max}}\\ 
\cline{2-7}
\bf metric ratio & \bf Data  & \bf Model & \bf Data  & \bf Model & \bf Data  & \bf Model \\ \thickhline
1.5~$\um$~:~3~$\um$ & 1.04  & 1.65 & 1.52  & 4.23 & 0.40  & 0.52 \\ \hline
1.5~$\um$~:~6~$\um$ & 4.16  & 4.92 & 5.00  & 20.7 & 0.26  & 0.09\tnote{\#1} \\ \hline
3~$\um$~:~6~$\um$   & 4.01  & 2.98 & 3.27  & 4.90 & 0.63  & 0.17\tnote{\#1} \\ \hline
\end{tabular}
\begin{tablenotes}
\item [\#1] For 6~$\um$ particles, the predicted concentration-time profile was very flat, resulting in a late  T\textsubscript{max} and therefore low 1.5~$\um$~:~6~$\um$ and 3~$\um$~:~6~$\um$ T\textsubscript{max} ratios.
\end{tablenotes}
\end{threeparttable}
\begin{flushleft} 
Comparison of model-predicted and reported PK between three different inhaled monodisperse particle formulations of fluticasone propionate, with aerodynamic diameters of 1.5, 3, and 6 $\um$, respectively \cite{usmani2005}. Due to uncertainty in reported absolute PK parameter readouts, the ratios between both listed particle sizes are reported instead. For example, the 1.5 $\um$-sized particles yielded a 4.16 fold higher measured $\text{AUC}_\text{0-12h}$ in comparison to the 6 $\um$-sized particles, whereas the model-based prediction resulted in 4.92 fold higher $\text{AUC}_\text{0-12h}$.  
\end{flushleft}
\end{table}

\begin{figure}[!h]
\centering
\includegraphics[width=\textwidth,trim = 20mm 40mm 20mm 30mm,clip]{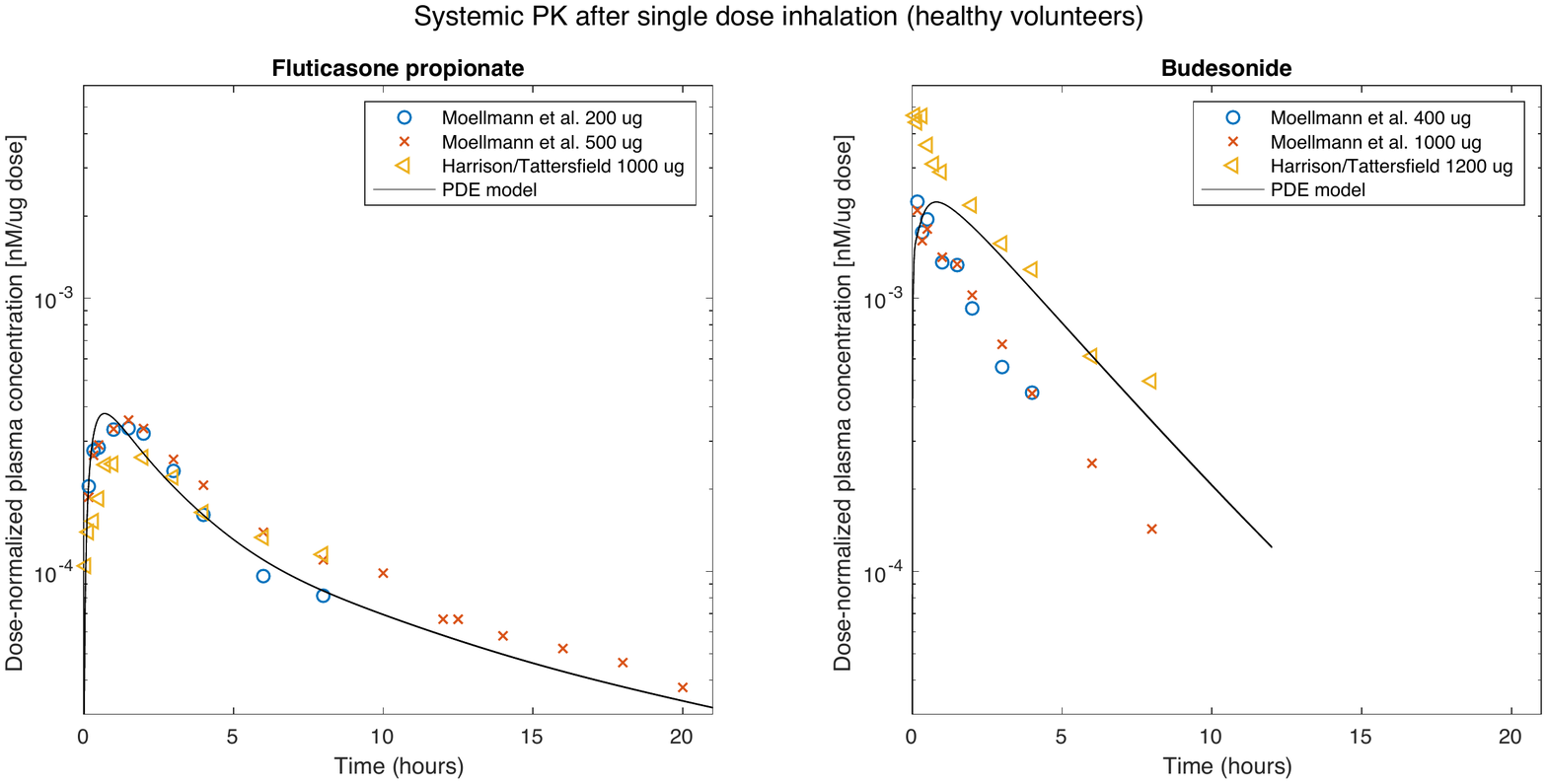}
\caption{{\bf Pharmacokinetics after drug inhalation of clinical formulations.}
Plasma concentration-time profiles after drug inhalation of fluticasone propionate inhaled with the \Diskus inhalation device (left panel) and budesonide inhaled with the \Turbohaler inhalation device (right panel). Data points: digitalized raw data from \cite{moellmann2001,harrison2003}, solid lines: PDE-model based predictions for 200, 500, and 1000 $\ug$ doses of fluticasone propionate and 400, 1000, and 1200 $\ug$ doses for budesonide (due to an almost dose-linear PK, model predictions overlap).
}
\label{lung_PK_healthy}
\end{figure}

As a last step of the PDE model evaluation, systemic PK profiles of fluticasone propionate and budesonide were simulated for both healthy volunteers and asthmatic patients, the only assumed difference between both populations being a more central particle deposition in asthma patients (see deposition profiles in \nameref{S2_Fig}). 
For fluticasone propionate inhaled by healthy volunteers with the \Diskus device, a C\textsubscript{max} of 0.38~nM per mg~dose, an AUC\textsubscript{0-12h} of $1.8~\text{nM}\!\cdot\!\text{h}$ per mg~dose, and a T\textsubscript{max} after 41~min were predicted. 
For budesonide (\Turbohaler), dissolution as well as absorption to the systemic circulation were predicted to be faster compared to fluticasone propionate, with a C\textsubscript{max} of 2.2~nM per mg~dose; T\textsubscript{max} was similar and AUC\textsubscript{0-12h} larger ($10~\text{nM}\!\cdot\!\text{h}$ per mg~dose).
The comparison of model-predicted PK profiles for fluticasone propionate and budesonide in comparison to observed clinical data from healthy volunteers \cite{moellmann2001,harrison2003} are displayed in Fig~\ref{lung_PK_healthy}. For fluticasone propionate, the dose-normalized data from literature were in agreement, and the simulation results closely matched these data. For budesonide, there was a between-study, but not within-study discrepancy between reported dose-normalized concentration-time profiles; model predictions were well within the reported range. The discrepancy in the data was not explainable by dose-nonlinear PK, since a 2.5-fold dose change in \cite{moellmann2001} did not impact on the normalized profiles. Of note, the model-predicted dose-normalized concentration-time profiles based on these scenarios all overlapped, which agrees with the clinically observed absence of dose-dependent pharmacokinetics in plasma for both fluticasone propionate and budesonide.

The same simulations for asthmatic patients resulted in lower systemic exposure. For fluticasone propionate, 28\% of the initially deposited lung dose was predicted to be eliminated via mucociliary clearance in healthy volunteers, compared to  53\% in asthmatic patients due to the more central particle deposition. For budesonide, 6\% and 29\% of the initially deposited lung dose were predicted to be eliminated via mucociliary clearance in healthy volunteers and asthmatic patients, respectively. A comparison of model-predicted and clinically observed differences between healthy volunteers and asthmatic patients is given in Table~\ref{healthy_vs_asthma}. For fluticasone propionate, simulations were in good agreement with clinical data, whereas for budesonide, the effect of the disease was overpredicted. However, the model-predicted stronger disease effect for fluticasone propionate compared to budesonide --in terms of an AUC increase-- was in agreement with the clinical data.

\begin{table}[!ht]
\centering
\begin{threeparttable}[b]
\caption{\label{healthy_vs_asthma}
{\bf Evaluation of model-predicted PK differences between healthy volunteers and asthmatic patients.}}
\begin{tabular}{|r+c|c|c|c|c|c|}
\hline
\bf Healthy:asthmatic  & \multicolumn{2}{|c|}{\bf AUC\textsubscript{0-12h}} & \multicolumn{2}{|c|}{\bf  C\textsubscript{max}}& \multicolumn{2}{|c|}{\bf  T\textsubscript{max}}\\ 
\cline{2-7}
\bf ratio for substance & \bf Data  & \bf Model & \bf Data  & \bf Model & \bf Data  & \bf Model \\ \thickhline
Fluticasone propionate & 1.76  & 1.64 & 1.67  & 1.48 & 1  & 1.05 \\ \hline
Budesonide & 0.88  & 1.43 & 1.07  & 1.51 & NA\tnote{\#1}  & 1.05 \\ \hline
\end{tabular}
\begin{tablenotes}
\item [\#1] since the reported T\textsubscript{max} values both corresponded to the first observed time point, no meaningful statement about T\textsubscript{max} ratios can be made.
\end{tablenotes}
\end{threeparttable}

\begin{flushleft} 
Comparison of model-based and literature-reported PK difference between healthy volunteers and asthmatic patients. Data are taken from \cite{harrison2003} (1000~$\ug$ fluticasone propionate with \Diskus / 1200~$\ug$ budesonide with \Turbohaler).  Ratios larger than 1 indicate higher values in healthy volunteers, whereas ratios smaller than 1 indicate higher values in asthmatic patients. NA, not available. 
\end{flushleft}
\end{table}

\subsection*{Sensitivity analysis}

\begin{figure}[!h]
\centering
\includegraphics[width=\textwidth,trim=0 30mm 0 30mm,clip]{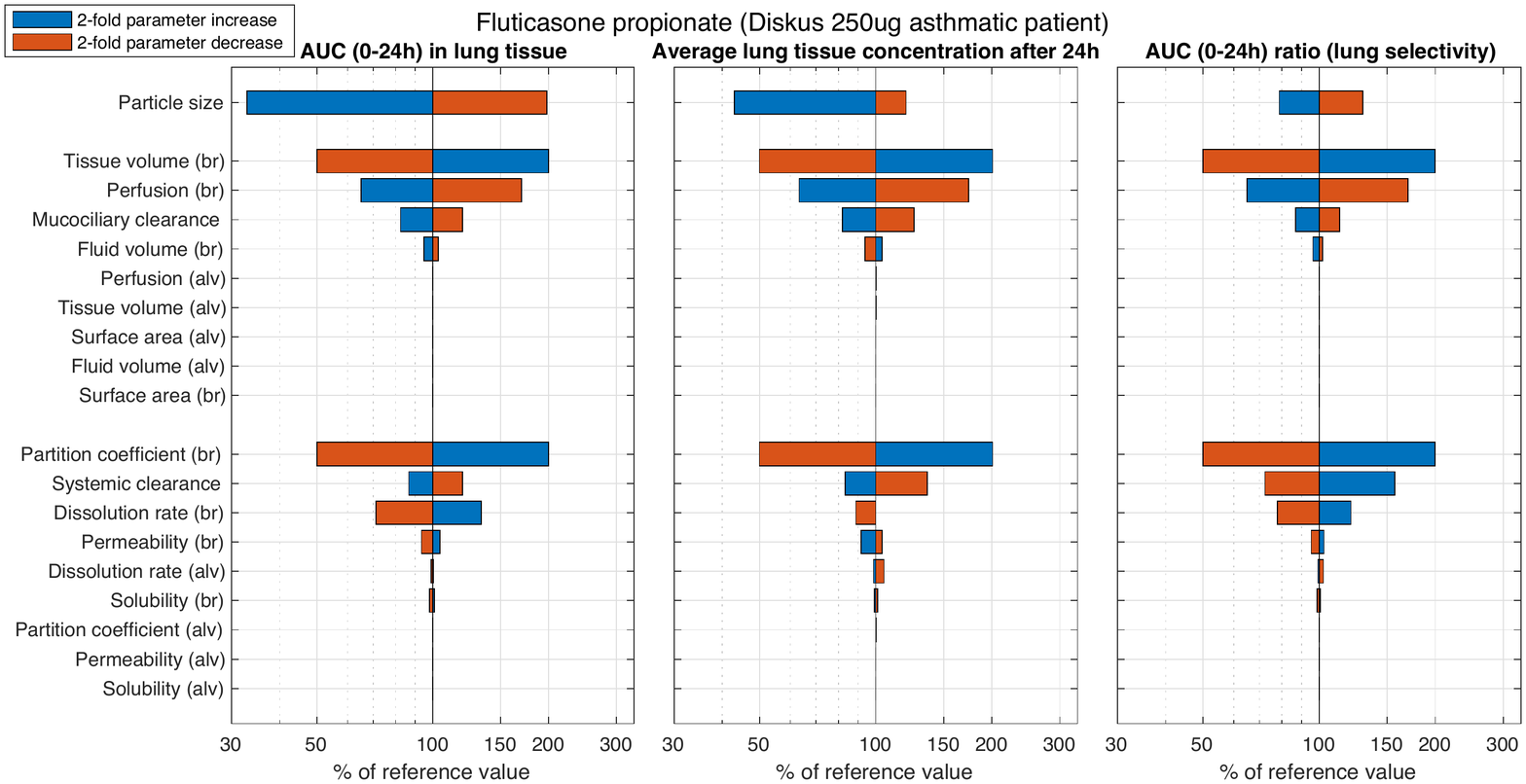}
\caption{\label{sensitivities}
{\bf Results of the performed sensitivity analysis for fluticasone propionate.}
For each of three different exposure measures readouts (AUC, $C_{24}$, and lung selectivity), the impact of a 2-fold increase (blue) and decrease (red) are depicted for a the formulation parameter particle size (top bar) and a set of physiological (middle bars) and drug-dependent parameters (bottom bars). The larger a bar, the stronger the impact of the varied parameter on the respective PK readout.
}
\end{figure}

As a last step of the analysis, a sensitivity analysis was applied to the evaluated PDE model to determine the most impactful parameters (among formulation-dependent, physiological, and drug-specific parameters) on the following PK readouts:
\begin{itemize}
\item[(i)]  AUC\textsubscript{0-24h} in conducting airway tissue,
\item[(ii)] the average concentration in the conducting airway tissues after 24~h (which is supposed to correlate with long-lasting efficacy of an inhaled drug), and 
\item[(iii)] lung selectivity, which is expressed as a ratio between the pulmonary AUC (in conducting airways) and the systemic AUC.
\end{itemize}  
This last quantity is supposed to provide a metric of local efficacy weighed against systemic safety, which is an important optimization criterion for inhaled drugs. As the relevance of an input parameter can depend on the complete set of the initial input parameters, the sensitivity analysis was performed starting with the parameters for (i) a 250~$\mu$g fluticasone propionate dose (see Fig.~\ref{sensitivities}) and (ii) a 800~$\mu$g budesonide dose (see \nameref{S1_Fig}), both representing approved doses \cite{diskus2000,turbohaler1998}.

Overall, the order of impactful parameters only differed marginally for the different exposure metrics and different drugs. A more than 50\% change was observed for tissue volume, tissue partition coefficient, perfusion, dissolution rate and systemic clearance. Particle size had a considerable impact for fluticasone propionate, and less for budesonide. 
For fluticasone propionate, the impact of drug solubility in the airway lining fluids was negligible, whereas for budesonide, although being the more soluble drug, a relevant impact was predicted since lining fluid concentrations approached the solubility (see \nameref{S3_Fig}).
Other parameters, such as lining fluid volume and all physiological parameters related to the alveolar space were characterized by negligible impact on the exposure metrics. The impact of deviating the model parameters by a 2-fold increase and 2-fold decrease was typically antithetical. As a notable exception, the dissolution rate in the conducting airways resulted in lower lung tissue concentration after 24~h, regardless of whether the dissolution rate was increased or decreased 2-fold. 

\section*{Discussion}

The pulmonary pharmacokinetics of orally inhaled drugs are highly complex as pulmonary deposition, pulmonary dissolution, mucociliary clearance, and pulmonary absorption create a complex interplay. Consequently, defining adequate optimization parameters for orally inhaled drugs remains challenging. To adequately capture and mechanistically predict the complex interplay of all pulmonary PK processes and to identify optimization parameters, a PDE-based mechanistic PK framework was developed. 

To build sufficient trust into a pulmonary PK model to use it for identification of optimal compound characteristics, an adequate and systematic model evaluation is a prerequisite. However, previous mechanistic modeling attempts, most noticeably the ones by Caniga et al. \cite{caniga2016} and Boger et al. \cite{boger2018}, lack such a thorough evaluation. Indeed, the approach by Caniga et al., differentiating between airways and alveolar space albeit less mechanistically than in the here-presented model, was evaluated for inhaled mometasone \cite{caniga2016} and more recently for additional fast dissolving drugs (formoterol, salbutamol, and budesonide) \cite{cabal2016}. However, these drugs would not provide the same insights into the pulmonary interplay of deposition, mucociliary clearance, and dissolution as the slowly dissolving drug fluticasone propionate. An adequate prediction quality for healthy and diseased populations, different particle sizes, slowly dissolving drugs or even insoluble particles remains to be demonstrated. 

A PDE model published by Boger et al. mechanistically included all pulmonary PK processes \cite{boger2018}. However, this model was based on a hypothetical drug, and while most of the characteristics of this hypothetical drug can be considered reasonable, such as a $K_\text{p,lung}$ of 4.9 or the oral bioavailability of 20\%, other characteristics such as a molecular weight of 250~Da were not considered typical for inhaled drugs (a typical molecular weight for an inhaled drug was reported to be $\approx$370~Da \cite{ritchie2009}). More importantly, since a model evaluation based on a hypothetical drug is not feasible, no assessment of the model's predictive capacities was made.

Therefore, the here-presented model represents --to the best of our knowledge-- the first systematically evaluated and publicly available mechanistic pulmonary PK model. First, to evaluate the mechanistic implementation of the mucociliary clearance, model-based lung retention profiles were compared to the pulmonary retention of insoluble gold and polystyrene particles. 24~h particle retention was adequately predicted, whereas long-term retention was underpredicted. 
One potential explanation could be that the assumed inhaled volume was larger than described, resulting in more particle deposition in the alveolar space. 
As the inhaled volume was defined based on the described experimental setup and not measured in the study, a deviation from the assumption is possible.
Since drug deposited in the alveolar space is not cleared by the mucociliary clearance, this would result in higher long-term lung retention than predicted. 
This explanation is reasonable due to high overall high (inter-subject) variability in pulmonary deposition \cite{stahlhofen1981}. 
As pulmonary drug retention over 10~days should not be relevant for orally inhaled drugs, however, this discrepancy was considered acceptable. 

Second, to evaluate the mechanistic implementation of the interplay of particle deposition, mucociliary clearance, and pulmonary drug dissolution, the PK of fluticasone propionate for different monodisperse particles (1.5, 3, and 6 $\um$ aerodynamic diameter) were predicted and compared to published data \cite{usmani2014}. It has to be stated that the reported absolute exposure metrics could not be reproduced. However, they appear extraordinarily high and could not be reached even if the provided dose had been administered intravenously. Nevertheless, the publication by Usmani et al.~contains a unique data set, and therefore we still considered the dataset, but by comparing the relative, not absolute, differences between the predictions for different particle sizes. Based on this evaluation, we considered the model-based predictions for the varying particle size effect as good.

Third, the modeling framework was used (without estimating additional input parameters) to simultaneously predict the PK of both fluticasone propionate and budesonide. For both drugs, plasma concentration-time profiles in healthy volunteers were very well predicted. In addition, the difference in pharmacokinetics between healthy volunteers and asthmatic patients was well predicted for fluticasone propionate. In contrast, the impact of disease on the PK of budesonide was overpredicted, i.e. in asthmatic patients more drug was predicted to be cleared by mucociliary clearance before it could be absorbed. This can be attributed to the strongly increased deposition of drug particles in the first airway generations (see \nameref{S2_Fig}), a prediction based on the assumption that the deposition probability across all airway generations is increased to a similar extent by local airway obstructions. In contrast, it was discussed that airway obstructions in asthma are located more peripherally in the conducting airways (in higher airway generations) \cite{deepak2017} and therefore the deposition would increase in more peripheral conducting airways rather than in the trachea and first airway generations (as can be seen in the imaging data in \cite{usmani2005}). Unfortunately, we are not aware of quantitative data or deposition models based on such data, which would allow to better account for these differences between healthy volunteers and asthmatic patients. Therefore, we were unable to integrate a more adequate representation into our mechanistic model. 

Based on the overall good agreement between the predictions and observed clinical data, we consider the here-published PDE-based PK model as the currently best-evaluated mechanistic model for orally inhaled drugs. However, even this mechanistic PK model still represents a simplification of reality and only includes the above-mentioned pulmonary PK processes; macrophage clearance as well as pulmonary metabolism were assumed not relevant. For some specific inhaled drugs, this assumption might not hold true. For example, pulmonary metabolism was discussed to be of importance for inhaled macromolecules (e.g., insulin \cite{sakagami2004,shen1999}). Macrophage clearance from the alveolar space to the conducting airways was characterized by a very long half-life of 35 -- 115 days  \cite{icrp1994,ncrp1997}. Consequently, compared to pulmonary absorption and dissolution kinetics of most inhaled drugs, macrophage clearance is expected to be negligible. Furthermore, the considerable between-study variability in reported data has to be kept in mind when judging the model evaluation accuracy. To recognize all of these assumptions, to understand their potential impact on the pulmonary PK, and finally to adequately apply the here presented model framework, a sound understanding of respiratory drug delivery remains essential.

As a last step of the presented analysis, we investigated the most relevant optimization parameters for orally inhaled drugs. To this end, we performed a model-based sensitivity analysis to identify the most impactful model parameters on pulmonary exposure metrics. The  pulmonary AUC was considered as a surrogate for pulmonary efficacy and the average concentration in the conducting airways after 24~h was considered a surrogate for the effect duration of an inhaled drug. Finally yet importantly, the ratio between pulmonary and systemic exposure was considered as a surrogate for lung selectivity of an inhaled drug (i.e. the larger the ratio, the better the lung selectivity). 

An impactful formulation-dependent model parameter was the particle size distribution of the inhaled fluticasone propionate formulation. This might not be surprising as the particle size simultaneously affects various pulmonary PK processes, i.e., larger particles deposit more centrally, dissolve slower and therefore a higher fraction of drug would be cleared by the mucociliary clearance. As a result, model-based predictions for larger particles indicated less lung exposure, shorter drug residence times in the lung, as well as a lower lung selectivity. In contrast, smaller fluticasone propionate particles would improve all exposure metrics. In conclusion, the model-based prediction framework indicates that reducing the particle size for inhaled fluticasone propionate would be a reasonable optimization parameter. However, this optimization parameter was predicted relevant only for fluticasone propionate. In contrast, the sensitivity analysis predicted no relevant impact of the particle size to be expected for a drug like budesonide. 

Impactful drug-specific optimization parameters for both drugs were (i) the lung partition coefficient, (ii) the systemic clearance, and (iii) the dissolution rate. An increase in the pulmonary partition coefficient, which indicates an increase in the pulmonary tissue affinity, was already previously suggested as an optimization parameter for lung selectivity \cite{backstrom2016a,borghardt2016,stocks2014}. This parameter however has to be considered carefully as a high tissue affinity / binding also would decrease the free pulmonary concentration. The systemic clearance had low impact on the pulmonary drug concentrations, but a higher systemic clearance provided a better lung selectivity. Therefore, especially for drugs with a critical systemic safety profile increasing the systemic clearance can be considered meaningful. In agreement, the relevance of a high systemic clearance to reduce systemic adverse effects for orally inhaled drugs was previously discussed \cite{borghardt2016,hochhaus1997}. The pulmonary dissolution rate for fluticasone propionate already seems to be nearly optimal to achieve a long-lasting efficacy, which would be a good property for a once-daily administered drug. An additional decrease in the dissolution kinetics was predicted to rather decrease the long-lasting efficacy. This finding is in agreement with recent observations that increasing the tissue affinity might be a better strategy to prolong the efficacy than slow dissolution \cite{begg2019}. Interestingly, while the dissolution rate constant can still be considered an optimization criterion, the solubility in the airway lining fluid was not impactful for fluticasone propionate. This underlines that actually the dissolution rate and not the solubility is important for pulmonary drug administration. For budesonide, which is characterized by faster dissolution kinetics compared to fluticasone propionate, the solubility was as important as the dissolution rate constant. The reason is that for budesonide, four parameters simultaneously increased local drug concentrations in the epithelial lining fluids: (i) a higher inhaled dose compared to fluticasone propionate, (ii) a higher fraction of the drug deposited in the lungs, (iii) a lower permeability of budesonide resulting in a higher residence time of dissolved drug, as well as (iv) a faster dissolution, which leads to more dissolved drug in the lining fluids.

Even though this sensitivity analysis provides good insights into potential optimization parameters, it has to be recognized that varying a single input parameter at a time might not always be realistic. For example, a higher lipophilicity would result in slower dissolution kinetics, higher permeability, and higher tissue affinity. Therefore, as an extension of the here presented sensitivity analysis, a multi-parameter investigation might be meaningful during compound optimization. Alternatively, the model-based evaluation allows comparing completely different drugs in a drug optimization program to select the best drug candidate. However, here we evaluated the impact of the input parameters on the exposure in the conducting airways. These exposure metrics only represent surrogate parameter and have to be carefully selected based on the mode of action and the target location, i.e., for a target that would be located in the alveolar space other exposure metrics should be considered relevant for a sensitivity analysis.

In addition to identifying optimization parameters, this sensitivity analysis allows addressing a second aspect, namely to identify the most impactful (physiological) model parameters which have to be understood to adequately predict the PK after oral inhalation. \emph{Vice versa}, not knowing the exact values of less impactful (physiological) parameters is less critical to predict the drug exposure in human. The most impactful physiological parameters were tissue volume, perfusion, and mucociliary clearance. Less important physiological parameters were, for example, fluid volume or surface area. An additional highly uncertain parameter was the more central deposition pattern for asthmatic patients (these were corrected with an empirical correction factor). Therefore, to improve the PK predictions for patients, it would be valuable to generate and implement quantitative lung imaging data in patients \cite{boger2015}. Another important uncertainty was the dissolution rate constant in the mucus. To our knowledge, no head-to-head comparison is available for \emph{in vivo} relevant dissolution assays for both dissolution in the mucus and the alveolar lining fluids. This was why we had to make an assumption, namely a fivefold slower dissolution in the conducting airways compared to the alveolar lining fluid. The reason for these adapted dissolution rate constants is that the epithelial lining fluid in the conducting airways --the mucus-- contains a lower concentration of surfactants, which facilitate dissolution \cite{wiedmann2000}, compared to the alveolar lining fluid. In addition, the upper layer of the mucus is characterized by a higher viscosity \cite{lai2009}, which can also lead to a slower dissolution in comparison to the alveolar space. However, even though this assumption described the data well, it should be verified with in vitro dissolution experiments. In contrast, other uncertain (physiological) input parameters, such as the volume of the lung lining fluids, were not impactful and therefore could be considered less critical. 

The previously mentioned data-based limitations also represent the main opportunities to improve the mechanistic PK model. First, it would significantly improve the applicability of the PK model framework if an adequate pulmonary deposition model for asthmatics was also implemented (and later also for e.g., idiopathic pulmonary fibrosis). Furthermore, a more mechanistic representation of tissue distribution (e.g., separating extra- vs. intracellular concentrations) might increase the predictive power for drugs with a high pulmonary tissue binding. Adapting the model to clinical PK data (e.g., by estimating parameters) might improve the description of clinical data, but this would normally not be feasible during compound optimization. Therefore, no pulmonary PK parameters were estimated in this work.

In conclusion, a PDE-based fully mechanistic pulmonary PK model was developed to perform model-based predictions of the pulmonary and systemic pharmacokinetics of orally inhaled drugs based on \emph{in vitro}, formulation-specific, drug-specific, as well as physiological data. To our knowledge, this model is the first fully mechanistic and systematically evaluated  pulmonary PK model. We also have shown that due to a large inter-study variability, model evaluation based on single (clinical) studies should be considered cautiously. This evaluated PK framework was applied to provide unique insights into optimization criteria for orally inhaled drugs by applying a model-based sensitivity analysis. It also provided insights which uncertainties of the modeling framework still can be improved. Overall, our analysis demonstrated that the model-based framework offers the potential to increase the quantitative understanding about inhaled drugs and ultimately, the model-based approach is applicable to optimizing drugs and formulations for inhalation therapy. 

\section*{Supporting information}

\paragraph*{S1 Appendix.}
\label{S1_Appendix}
{\bf Model development and evaluation details.} Derivations of model components from first principles, a description of the numerical resolution method, model evaluation against nonclinical data and the strategy for deposition adaptation for asthmatic patients.

\paragraph*{S1 Fig.}
\label{S1_Fig}
{\bf Sensitivity analysis for budesonide.} Additional sensitivity analyses for budesonide.

\paragraph*{S2 Fig.}
\label{S2_Fig}
{\bf Deposition patterns.} Deposition patterns in healthy volunteers and asthmatic patients for fluticasone propionate (\Diskus) and budesonide (\Turbohaler).

\paragraph*{S3 Fig.}
\label{S3_Fig}
{\bf Concentration in lung lining fluids.} Predicted time- and location-resolved lung lining fluid concentrations of fluticasone propionate (\Diskus, 250 $\mu$g dose) and budesonide (\Turbohaler, 800 $\mu$g dose).

\paragraph*{S1 File.}
\label{S1_File}
{\bf MATLAB implementation.} MATLAB scripts used for solving the PDE model and for evaluating it against clinical and experimental data.

\paragraph*{S1 Table.}
\label{S1_Table}
{\bf Summary of pharmacokinetic studies.} All identified clinical studies on fluticasone propionate and budesonide, including their study design and reported pharmacokinetic parameters.

\section*{Acknowledgments}
We thank Carmen Hummel and the In Vitro ADME and CMC groups from Boehringer~Ingelheim~Pharma~GmbH~\&~Co.~KG for generating the \emph{in vitro} measurements reported as ``in-house data'' in the manuscript. Furthermore, we would like to thank the working group Mathematical Modelling and Systems Biology at the Institute of Mathematics of University of Potsdam, as well as colleagues from Boehringer~Ingelheim~Pharma~GmbH~\&~Co.~KG for proof-reading the manuscript.

\end{document}


\title{A mechanistic framework for \emph{a priori} pharmacokinetic predictions of orally inhaled drugs\\[5mm] S1~Appendix}

\maketitle 

\tableofcontents

\clearpage
\section{Derivation of the PDE model}

\subsection{Derivation of the dissolution model}

The Noyes-Whitney equation \cite{noyes1897} describes the dissolution flux $\ddt{W}$ in terms of properties of the dissolving particles and the dissolution medium,

\begin{equation} 
\label{noyeswhitney}
\ddt{W} = -\frac{D \cdot \text{SA}}{h}(C_\text{s}-C_\text{flu}),
\end{equation}
where $D$ is particle diffusivity, $\text{SA}$ particle surface area, $h$ height of the diffusion layer, $C_s$ particle solubility and $C_\text{flu}$ concentration of dissolved substance in the medium.

Through geometric assumptions on particles, this equation can be turned into a differential equation describing the change of volume of a dissolving particle. We assume particles to be spherical in shape, with radius $r$, surface area $\text{SA} = 4\pi r^2$, volume $s=\cdot\frac{4}{3}\pi r^3$ and mass $W = \rho s$. Furthermore, as suggested previously \cite{boger2016}, we assume the height of the diffusion layer to equate particle radius, $h\approx r$. 

Since parametrizing the model in terms of radius $r$ leads to a singularity of the dissolution model when $r\searrow 0$, in contrast to \cite{boger2016} we choose particle volume $s=\cdot\frac{4}{3}\pi r^3$ as a size descriptor instead of particle radius.

Differentiating the particle mass equation, 
\begin{equation}
\label{massflow}
\ddt{W}(t) = \rho \ddt{s}(t),
\end{equation}
and equating Eqs.~\eqref{noyeswhitney} and \eqref{massflow} yields

\[\ddt{s}(t) = -\frac{D \cdot 4\pi r(t)^2}{\rho\, r(t)}(C_\text{s}-C_\text{flu}) = -\frac{D \cdot 4\pi r(t)}{\rho}(C_\text{s}-C_\text{flu}) = -\frac{D \cdot 4\pi \Big(\frac{s(t)}{\frac{4}{3}\pi}\Big)^{1/3}}{\rho}(C_\text{s}-C_\text{flu}).\]
We opt to parametrize the dissolution model in terms of maximum dissolution rate $k_\text{diss}= D\cdot C_s$ rather than diffusivity $D$, since dissolution rate can be identified more directly from in vitro experiments (see Section \nameref{rohrschneider}). The resulting dissolution model reads

\[d(s,C_\text{flu}) = \frac{4\pi \,k_\text{diss}}{(\frac{4}{3}\pi)^{1/3}\,\rho}\cdot \left(1-\frac{C_\text{flu}}{C_s}\right) \cdot s^{1/3},\qquad \ddt{s}(t) = -d\big(s(t),C_\text{flu}\big).\]
The concentration of dissolved substance, $C_\text{flu}$, also changes during dissolution. These processes are coupled in the PDE model described below.

\subsection{Derivation of the mucociliary clearance model}

As explained in the main text, a continuous representation of airway radius $r(x)$ depending on location $x$ within the conducting airways is derived by interpolation. Using the Hofmann/Sturm model
\[v = 0.12553 \frac{\text{cm}}{\text{min}} \cdot \left(\frac{d}{1~\text{cm}}\right)^{2.808},\]
we obtain a location-dependent mucociliary clearance model for a particle at location $x(t)$ at time $t$:
\[\ddt{x}(t) = -\lambda_\text{mc}\big(x(t)\big) = 0.12553 \frac{\text{cm}}{\text{min}} \cdot \left[2r^\text{br}\left(\frac{x(t)}{1~\text{cm}}\right)\right]^{2.808} = -0.8791 \frac{\text{cm}}{\text{min}}\cdot r^\text{br}\left(\frac{x(t)}{1~\text{cm}}\right)^{2.808} 
.\]

\subsection{Individual and population states}

Physiologically-structured models describe the time evolution of a set of individuals/particles, each exhaustively described by a vector of characteristics called \emph{state}, denoted $z$, and which changes over time. The time evolution of the state of any individual is assumed to be governed by a law $G$, i.e.
\[\ddt{z}(t) = G(t,z(t)),\quad z(0) = z_0.\]
Assuming that a population consists of a large number of individuals, it is natural not to describe each single individual but rather the\ time evolution of a density $\rho(t,z)$ of individuals over the state space. In this representation, the total number of particles is given by
\[N(t) = \int \rho(t,z)\text{d}z,\]
and the number of particles within a particular subregion $\omega$ of the state space is given by
\[N_\omega(t) = \int_\omega \rho(t,z)\text{d}z.\]
For such a domain $\omega$, we set $\omega(t) = \{ z(t): z_0\in\omega\}$.
Assuming that the number of individuals is conserved in the state space, we obtain
\[\ddt{}N_{\omega(t)}(t) \equiv 0\]
as long as $\omega(t)$ does not touch the state space boundary. From this expression, a so-called continuity equation can be derived (see \cite{metz1986}):
\begin{equation}
\label{continuity}
\partial_t\rho(t,z) + \text{div}_z\big[G(t,z)\rho(t,z)\big] = 0.
\end{equation}

\subsection{Derivation of physiologically-structured population models (PSPMs)}

In our application context, the population consists of inhaled undissolved drug particles of different sizes, deposited at different locations within the conducting airways or within the alveolar space.
The number of particles can only change if particles are (i) cleared to the GI tract by the mucociliary elevator (mucociliary clearance beyond the trachea, $x(t)=0$) or
(ii) completely dissolved ($s(t)=0$).

\subsubsection{Conducting airways}

\begin{figure}[ht!]
\begin{center}
\includegraphics[width=.5\textwidth, trim = 67 60 47 60, clip]{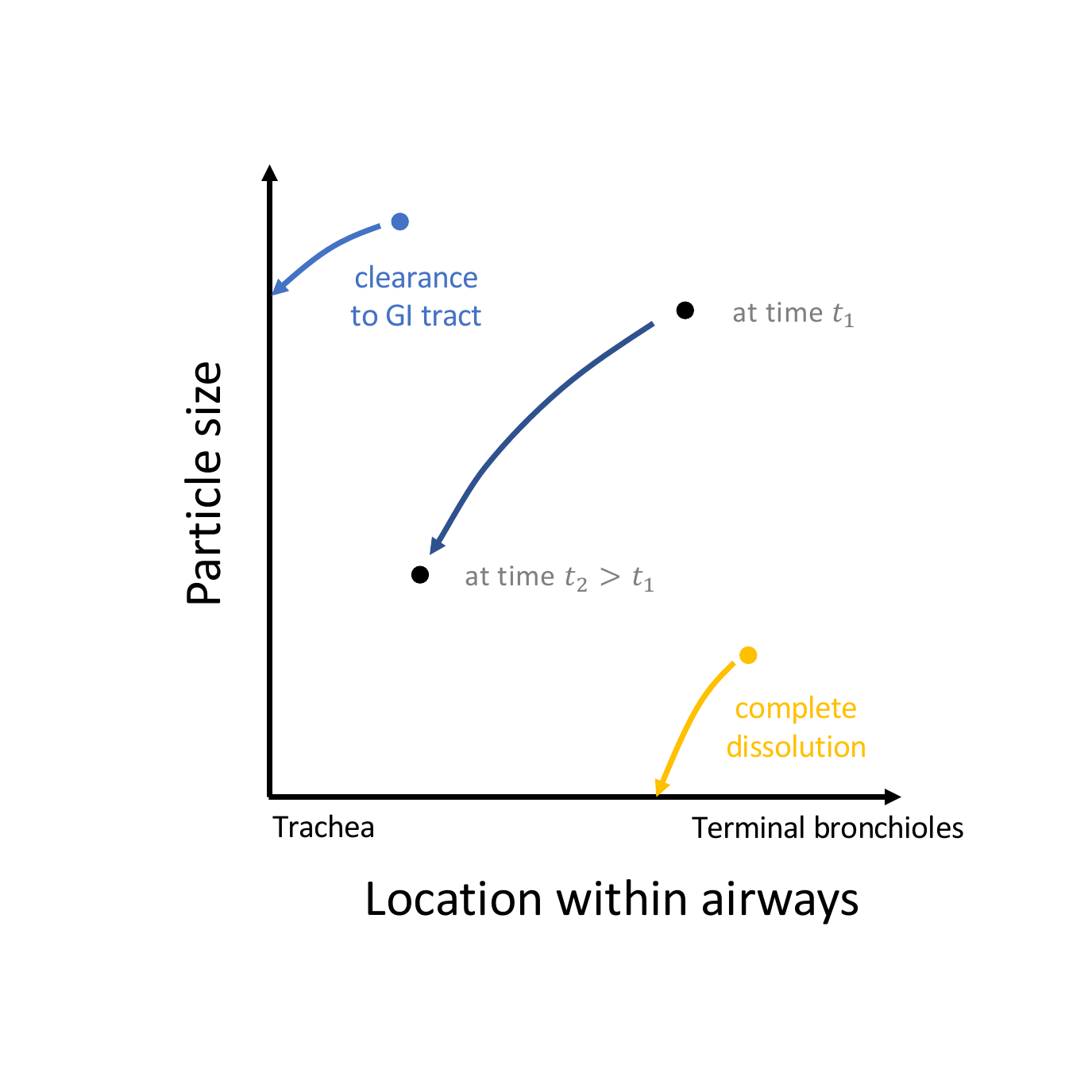}
\end{center}
\caption{ \label{fig:phaseplane} Phase plane representation of a drug particle in the conducting airways. Each particle is characterized by its location and size. Over time, particles move within this two-coordinate system until they are either cleared to the GI tract or completely dissolved.}
\end{figure}

The particle state $z=(x,s)\in[0,x_\text{TB}]\times[0,s_\text{max}]$ can change by mucociliary clearance or dissolution (illustrated in Fig~\ref{fig:phaseplane}):
\[\begin{pmatrix} \ddt{x}(t)\\ \ddt{s}(t)
\end{pmatrix} = 
\underbrace{\begin{pmatrix}
- \lambda_\text{mcc}(x(t))\\
-d(s(t),C^\text{br}_\text{flu}(x(t),t))
\end{pmatrix}}_{=:G^\text{br}(t,x(t),s(t))},\]
and Eq.~\eqref{continuity} yields the location- and size-structured bronchial PSPM 
\begin{equation}
\label{eq:pspm-br}
\partial_t\rho^\text{br}(t,x,s) - \partial_x \Big[\lambda_\text{mcc}(x) \rho^\text{br}(t,x,s)\Big] - \partial_s \Big[d\big(s,C^\text{br}_\text{flu}(t,x)\big) \rho^\text{br}(t,x,s)\Big] = 0.
\end{equation}

\subsubsection{Alveolar space}

Since mucociliary clearance is not present in the alveolar space, the particle state $z=s\in[0,s_\text{max}]$ can change by dissolution only:
\[\ddt{s}(t) = \underbrace{-d(s(t),C^\text{alv}_\text{flu}(t))}_{=:G^\text{alv}(t,s(t))},\]
and Eq.~\eqref{continuity} yields the size-structured alveolar PSPM 
\[\partial_t\rho^\text{alv}(t,s) - \partial_s \Big[d\big(s,C^\text{alv}_\text{flu}(t)\big) \rho^\text{alv}(t,s)\Big] = 0.\]

\subsection{Mass balances}

When coupling the PSPM models to equations for dissolved drug in lining fluids, the number of molecules (not particles) have to be conserved during dissolution and mucociliary clearance. This model feature is ensured by deriving dissolution and mucociliary clearance rates directly from the PSPMs, which is shown in the following. The number of undissolved molecules in the conducting airways / the alveolar space are given by
\[
A^\text{br}_\text{sol}(t)  = \int\limits_{0}^{x_\text{TB}}\int\limits_0^{s_\text{max}} s\rho^\text{br}(t,x,s) \text{d}x\text{d}s,\qquad
A^\text{alv}_\text{sol}(t) = \int\limits_0^{s_\text{max}} s\rho^\text{alv}(t,s)\text{d}s.
\]
We illustrate the derivation for the conducting airways, using integration by parts at step $(\ast)$:

\begin{align*}
\ddt{A^\text{br}_\text{sol}}(t) &= \int\limits_{0}^{x_\text{TB}}\int\limits_0^{s_\text{max}} s\partial_t\rho^\text{br}(t,x,s) \text{d}x\text{d}s\\
&\stackrel{\eqref{eq:pspm-br}}{=} -\int\limits_{0}^{x_\text{TB}}\int\limits_0^{s_\text{max}} s\Big(-\partial_x[\lambda_\text{mcc}(x)\rho^\text{br}(t,x,s)] - \partial_s[d(s,C_\text{flu}^\text{br}(t,x))\rho^\text{br}(t,x,s)]\Big)\text{d}x\text{d}s\\
&\stackrel{(\ast)}{=} \int\limits_0^{s_\text{max}} s \big(\lambda_\text{mcc}(x_\text{TB})\underbrace{\rho^\text{br}(t,\,x_\text{TB},\,s)}_{=0 \text{ (no inflow)}}-\lambda_\text{mcc}(0)\rho^\text{br}(t,0,s) \big)\text{d}s \\
&\quad- \int\limits_{0}^{x_\text{TB}}\int\limits_0^{s_\text{max}} d\big(s,C_\text{flu}^\text{br}(t,x)\big)\rho^\text{br}(t,x,s)\text{d}x\text{d}s +
\int\limits_{0}^{x_\text{TB}} s_\text{max} d\big(s_\text{max},C_\text{flu}^\text{br}(t,x)\big)\underbrace{\rho^\text{br}(t,x,s_\text{max})}_{=0 \text{ (no inflow)}} \text{d}x\\
&= -\underbrace{\int\limits_0^{s_\text{max}} s \,\lambda_\text{mcc}(0)\rho^\text{br}(t,0,s) \text{d}s}_\text{cleared by mucociliary elevator}
- \underbrace{\int\limits_{0}^{x_\text{TB}}\int\limits_0^{s_\text{max}} d\big(s,C_\text{flu}^\text{br}(t,x)\big)\rho^\text{br}(t,x,s)\text{d}x\text{d}s}_\text{dissolved into lining fluids}\\
\end{align*}

A similar but simplified reasoning applies to the alveolar space, where only dissolution, not mucociliary clearance, needs to be considered.

\clearpage
\section{Numerical resolution of the PDE model}

\subsection{Notation}

We consider a uniform time discretization step $\Delta t>0$, a location discretization 
\[0=x_{1/2}<...<x_{K+1/2}=x_\text{TB}\]
and a size discretisation
\[0=s_{1/2}<...<s_{L+1/2}=s_\text{max}\]

These discretization points are understood as vertices of mesh elements $(k,l)=[x_{k-1/2},x_{k+1/2}]\times [s_{l-1/2},s_{l+1/2}]$ within which unknowns (approximations of $\rho^\text{br}$, $C^\text{br}_\text{flu}$, etc.) are defined; they appear in the discretization of the location- and size-structured model in the conducting airways. The same size grid is also used when discretizing the size-structured model in the alveolar space. Furthermore, we define the center $(x_k,s_l)$ of mesh element $(k,l)$ from the above discretization points,
\begin{align*}
&&x_k &:= \frac{x_{k-1/2}+x_{k+1/2}}{2}, && k\in\{1,..,K\},&&\\
&&s_l &:= \frac{s_{l-1/2}+s_{l+1/2}}{2}, && l\in\{1,..,L\}.&&
\end{align*}

We use the following notation:
\begin{itemize}
\item $\Delta x_k:=x_{k+1/2}-x_{k-1/2}$ (location length of mesh element $(k,\cdot)$)
\item $\Delta s_l:=s_{l+1/2}-s_{l-1/2}$ (size length of mesh element $(\cdot,l)$); we also define $\Delta s_{l+1/2}:=s_{l+1}-s_{l}$ (this expression will appear later during computations)
\item Abbreviations for location-structured physiology in conducting airways: $\lambda_k:=\lambda_\text{mc}(x_k)$, $r_k^\text{br}:=r^\text{br}(x_k)$, $q_k^\text{br}:=q^\text{br}(x_k)$, $a^\text{br}_{\text{flu},k}:=a^\text{br}_\text{flu}(x_k)$, $a^\text{br}_{\text{tis},k}:=a^\text{br}_\text{tis}(x_k)$
\item $\rho_{k,l}^{\text{br},n}$ as the numerical approximation of $\rho^\text{br}(t_n,x_k,s_l)$ 
\item $\rho_{l}^{\text{alv},n}$ as the numerical approximation of $\rho^\text{alv}(t_n,s_l)$ 
\item $C^{\text{br},n}_{\text{flu},k}$ as the numerical approximation of $C^\text{br}_\text{flu}(t_n,x_k)$
\item $C^{\text{br},n}_{\text{tis},k}$ as the numerical approximation of $C^\text{br}_\text{tis}(t_n,x_k)$
\item $C^{\text{alv},n}_{\text{flu}}$ as the numerical approximation of $C^\text{alv}_\text{flu}(t_n)$
\item $C^{\text{alv},n}_{\text{tis}}$ as the numerical approximation of $C^\text{alv}_\text{tis}(t_n)$
\item $A^{\text{y},n}_x$ as the numerical approximation of $A^\text{y}_x(t_n)$ (total amount of drug in a certain state; one of 
$A^\text{br}_\text{sol}$, $A^\text{br}_\text{flu}$, $A^\text{br}_\text{tis}$, $A^\text{alv}_\text{sol}$, $A^\text{alv}_\text{flu}$, $A^\text{alv}_\text{tis}$, $A^\text{clear}_\text{mcc}$, $A^\text{clear}_\text{sys}$, $A^\text{sys}_\text{tot}$), with 'sol' meaning 'solid', i.e. undissolved.
\end{itemize}

\subsection{Upwind discretization of physiologically-structured population equations}

Upwind discretizations, i.e. non-centered finite difference approximations depending on the flow direction, are well tailored to PSPMs, resulting in stable discretizations as long as the timestep $\Delta t$ is small enough (called a CFL condition).

The upwind discretization of the conducting airway PSPM 
\[\partial_t\rho^\text{br}(t,x,s) - \partial_x\big[\lambda_\text{mc}(x)\rho^\text{br}(t,x,s)\big] - \partial_s\big[d(s,C_\text{flu}^\text{br}(t,x))\rho^\text{br}(t,x,s)\big] = 0\]
is given by
\[\frac{\rho_{k,l}^{\text{br},n+1}-\rho_{k,l}^{\text{br},n}}{\Delta t} - \frac{\lambda_{k+1/2}\rho_{k+1,l}^{\text{br},n} - \lambda_{k-1/2}\rho_{k,l}^{\text{br},n}}{\Delta x_k} - \frac{d(s_{l+1/2},C^{\text{br},n}_{\text{flu},k})\rho_{k,l+1}^{\text{br},n} - d(s_{l-1/2},C^{\text{br},n}_{\text{flu},k})\rho_{k,l}^{\text{br},n}}{\Delta s_l}=0,\]
for $n\in\{1,...,N\}, k\in\{1,...,K\}, l\in\{1,...,L\}$ (with $\rho_{K+1,l}^{\text{br},n}=\rho_{k,L+1}^{\text{br},n}=0$, i.e. no inflow condition). Similarly the upwind discretization of the alveolar PSPM 
\[\partial\rho^\text{alv}(t,s) - \partial_s\big[d(s,C^\text{alv}_\text{flu}(t))\rho^\text{alv}(t,s)\big] = 0\]
is given by
\begin{equation*}
\frac{\rho^\text{alv,n+1}_{l}-\rho^{\text{alv},n}_{l}}{\Delta t} - \frac{d(s_{l+1/2},C^{\text{alv},n}_\text{flu})\rho^{\text{alv},n}_{l+1} - d(s_{l-1/2},C^{\text{alv},n}_\text{flu})\rho^{\text{alv},n}_{l}}{\Delta s_l}=0.
\end{equation*}
Within this framework, the number of undissolved drug molecules is approximated by
\begin{align*}
A_\text{sol,k}^{\text{br},n} &:= \sum_{l=1}^L \Delta s_l s_l\rho_{k,l}^{\text{br},n} && \text{(location $k$ in conducting airways)},\\
A_\text{sol}^{\text{alv},n} &:= \sum_{l=1}^L \Delta s_l s_l\rho_{l}^{\text{alv},n} && \text{(alveolar space)}.\\
\end{align*}

\subsection{Implicit discretization of linear processes}

Recognizing that all processes except for dissolution and mucociliary clearance are linear, we propose an implicit discretization to ensure unconditional stability of these other processes, too. The numerical scheme is formulated in terms of local amounts (in bronchial/alveolar fluid/tissue) rather than concentrations. To this end, we define
\begin{align*}
V^\text{br}_{\text{flu},k}&:=\Delta x_k\, a^\text{br}_{\text{flu},k}\qquad \text{(lining fluid volume at $k$-th location grid cell)}\\
V^\text{br}_{\text{tis},k}&:=\Delta x_k\, a^\text{br}_{\text{tis},k}\qquad \text{(tissue volume at $k$-th location grid cell)}
\end{align*}
and obtain the amounts
\begin{align*}
A_{\text{flu},k}^{\text{br},n} &:= C_{\text{flu},k}^{\text{br},n} V^\text{br}_{\text{flu},k},  &&A_{\text{tis},k}^{\text{br},n} := C_{\text{tis},k}^{\text{br},n} V^\text{br}_{\text{tis},k} && \text{(conducting airways)},\\
A_{\text{flu}}^{\text{alv},n} &:= C_{\text{flu}}^{\text{alv},n} V_{\text{flu}}^\text{alv}, &&A_{\text{tis}}^{\text{alv},n} := C_{\text{tis}}^{\text{alv},n} V_{\text{tis}}^\text{alv} && \text{(alveolar space)}.
\end{align*}
Furthermore, it will be useful to define
\begin{align*}
\text{PS}^\text{br}_\text{k}&:=\Delta x_k \, 2\pi r_k^\text{br} \, P_\text{app} \quad \text{(permeability-surface area product at $k$-th location grid cell)}\\
Q^\text{br}_k&:=\Delta x_k\, q^\text{br}_k\quad\qquad\! \text{(perfusion of $k$-th location grid cell)}.
\end{align*}
To arrive at a numerical scheme formulated on the computational grid, integrals are discretized as follows:

\[\int_0^\text{s\textsubscript{max}} f(s)\text{d}s \quad\Rightarrow\quad \sum_{l=1}^L \Delta s_l f(s_l),\qquad\int_0^\text{x\textsubscript{TB}} f(x)\text{d}x \quad\Rightarrow\quad \sum_{k=1}^K \Delta x_k f(x_k)\]

\paragraph{Bronchial kinetics}
\begin{align*}
\frac{A^{\text{br},n+1}_{\text{flu},k}-A^{\text{br},n}_{\text{flu},k}}{\Delta t} &= \underbrace{
\Delta x_k\sum_{l=2}^{L} \Delta s_{l-1/2} d(s_{l-1/2},C^{\text{br},n}_{\text{flu},k}) \rho_{k,l}^{\text{br},n}}_\text{dissolved (see section below)}
- \;\text{PS}^\text{br}_\text{k} \left(\frac{A^{\text{br},n+1}_{\text{flu},k}}{V^\text{br}_{\text{flu},k}} - \frac{A^{\text{br},n+1}_{\text{tis},k}}{V^\text{br}_{\text{tis},k}\,K_{pl,u}}\right)\\
\frac{A^{\text{br},n+1}_{\text{tis},k}-A^{\text{br},n}_{\text{tis},k}}{\Delta t} &= 
\text{PS}^\text{br}_\text{k} \left(\frac{A^{\text{br},n+1}_{\text{flu},k}}{V^\text{br}_{\text{flu},k}} - \frac{A^{\text{br},n+1}_{\text{tis},k}}{V^\text{br}_{\text{tis},k}\,K_{pl,u}}\right)
 - Q^\text{br}_k \left(\frac{A^{\text{br},n+1}_{\text{tis},k}}{V^\text{br}_{\text{tis},k}} \frac{R}{K_{pl}}  - \frac{A_\text{ctr}^{\text{sys},n+1}}{V^\text{sys}_\text{ctr}}\right)
\end{align*}

\paragraph{Alveolar kinetics}
\begin{align*}
\frac{A^{\text{alv},n+1}_{\text{flu}}-A^{\text{alv},n}_{\text{flu}}}{\Delta t} &= 
\sum_{l=2}^{L} \Delta s_{l-1/2} d(s_{l-1/2},C^{\text{alv},n}_\text{flu}) \rho_{l}^{\text{alv},n}
- \text{PS}^\text{alv} \left(\frac{A^{\text{alv},n+1}_\text{flu}}{V^\text{alv}_\text{flu}} - \frac{A^{\text{alv},n+1}_{\text{tis}}}{V^\text{alv}_{\text{tis}}\,K_{pl,u}}\right)\\\frac{A^{\text{alv},n+1}_{\text{tis}}-A^{\text{alv},n}_{\text{tis}}}{\Delta t} &= 
\text{PS}^\text{alv} \left(\frac{A^{\text{alv},n+1}_\text{flu}}{V^\text{alv}_\text{flu}} - \frac{A^{\text{alv},n+1}_\text{tis}}{V^\text{alv}_\text{tis}\,K_{pl,u}}\right)
 - Q^\text{alv} \left(\frac{A^{\text{alv},n+1}_\text{tis}}{V^\text{alv}_\text{tis}} \frac{R}{K_{pl}}  - \frac{A_\text{ctr}^{\text{sys},n+1}}{V^\text{sys}_\text{ctr}}\right)
\end{align*}

\paragraph{Systemic kinetics}
\begin{align*}
\frac{A^{\text{sys},n+1}_{\text{gut}}-A^{\text{sys},n}_{\text{gut}}}{\Delta t} &= 
\underbrace{\sum_{l=1}^{L} \Delta s_{l} s_l \lambda_{1/2} \rho_{1,l}^{\text{br},n}}_\text{mucociliary clearance (see section below)}
- k_{01}A^{\text{sys},n+1}_{\text{gut}}\\
\frac{A^{\text{sys},n+1}_\text{ctr}-A^{\text{sys},n}_\text{ctr}}{\Delta t} &= Fk_{01}A^{\text{sys},n+1}_{\text{gut}} -k_{12}A^{\text{sys},n+1}_\text{ctr} + k_{21}A^{\text{sys},n+1}_\text{per}\\
& \quad+ Q^\text{alv} \left(\frac{A^{\text{alv},n+1}_\text{tis}}{V^\text{alv}_\text{tis}} \frac{R}{K_{pl}}  - \frac{A_\text{ctr}^{\text{sys},n+1}}{V^\text{sys}_\text{ctr}}\right)\\
& \quad+ \sum_{k=1}^K Q^\text{br}_k \left(\frac{A^{\text{br},n+1}_{\text{tis},k}}{V^\text{br}_{\text{tis},k}} \frac{R}{K_{pl}}  - \frac{A_\text{ctr}^{\text{sys},n+1}}{V^\text{sys}_\text{ctr}}\right)\\
\frac{A^{\text{sys},n+1}_\text{per}-A^{\text{sys},n}_\text{per}}{\Delta t} &= 
k_{12}A^{\text{sys},n+1}_\text{ctr} - k_{21}A^{\text{sys},n+1}_\text{per}\\
\frac{A^{n+1}_\text{clear} - A^{n}_\text{clear}}{\Delta t} &= 
(1-F)k_{01}A^{\text{sys},n+1}_{\text{gut}} + k_{10} A^{\text{sys},n+1}_\text{ctr}
\end{align*}

\subsection{Mass conservation of PDE discretisation}

The above terms are chosen such that the number of molecules is conserved, i.e., the total amount of drug in the body plus the amount excreted, given by
\[A_\text{tot}^n = \underbrace{A_\text{sol}^{\text{br},n} + \sum_{k=1}^K \Big(A^{\text{br},n}_\text{flu,k} + A^{\text{br},n}_\text{tis,k}\Big)}_\text{in conducting airways} + \underbrace{A_\text{sol}^{\text{alv},n} + A^{\text{alv},n}_\text{flu} + A^{\text{alv},n}_\text{tis}}_\text{in alveolar space} + \underbrace{A^{\text{sys},n}_\text{ctr} + A^{\text{sys},n}_\text{per} + A^{\text{sys},n}_\text{gut} + A^{n}_\text{clear}}_\text{in GI tract, systemic circulation or excreted},\]
remains constant for all $n$. 
Mass conservation during uptake from lining fluid to lung tissue can be seen directly from the rates in the equations: the same terms, e.g.
\[\text{PS}^\text{br}_\text{k} \left(\frac{A^{\text{br},n+1}_{\text{flu},k}}{V^\text{br}_{\text{flu},k}} - \frac{A^{\text{br},n+1}_{\text{tis},k}}{V^\text{br}_{\text{tis},k}\,K_{pl,u}}\right),\]
appear in both equations with opposing signs (for systemic uptake from conducting airway tissue, contributions at different locations are summed).  Furthermore, using the upwind formulation, we can decompose the rate of change of the amount of undissolved drug:
\begin{align*}
\frac{A_\text{sol}^{\text{br},n+1}-A_\text{sol}^{\text{br},n}}{\Delta t} &=
\sum_{k=1}^K\sum_{l=1}^L\Delta x_k \Delta s_l\, s_l\frac{\rho_{k,l}^{\text{br},n+1}-\rho_{k,l}^{\text{br},n}}{\Delta t}\\
&= - \sum_{k=1}^K\sum_{l=1}^L\Delta x_k \Delta s_l\, s_l \left(-\frac{\lambda_{k+1/2}\rho_{k+1,l}^{\text{br},n} - \lambda_{k-1/2}\rho_{k,l}^{\text{br},n}}{\Delta x_k} - \frac{d(s_{l+1/2},C^{\text{br},n}_{\text{flu},k})\rho_{k,l+1}^{\text{br},n} - d(s_{l-1/2},C^{\text{br},n}_{\text{flu},k})\rho_{k,l}^{\text{br},n}}{\Delta s_l}\right) \\
&= + \sum_{k=1}^K\sum_{l=1}^L\Delta s_l\, s_l \Big(\lambda_{k+1/2}\rho_{k+1,l}^{\text{br},n} - \lambda_{k-1/2}\rho_{k,l}^{\text{br},n}\Big)\\
&~\quad + \sum_{k=1}^K\sum_{l=1}^L\Delta x_k s_l \Big(d(s_{l+1/2},C^{\text{br},n}_{\text{flu},k})\rho_{k,l+1}^{\text{br},n} - d(s_{l-1/2},C^{\text{br},n}_{\text{flu},k})\rho_{k,l}^{\text{br},n}\Big)\\
 &= -\underbrace{\sum_{l=1}^L\Delta s_l\, s_l \lambda_{1/2}\rho_{1,l}^{\text{br},n}}_\text{mucociliary clearance}
 - \sum_{k=1}^K \underbrace{\Delta x_k\sum_{l=2}^{L} \Delta s_{l-1/2} d(s_{l-1/2},C^{\text{br},n}_{\text{flu},k}) \rho_{k,l}^{\text{br},n}}_\text{dissolution at location $k$}
\end{align*}
and noting that these two terms are matched in the equations for dissolved drug in the lining fluid 
and of cleared drug, we can conclude that mass is conserved during dissolution and mucociliary clearance. An analogous computation shows mass conservation during dissolution in the alveolar space. Mass balance was checked systematically during all simulations shown.

\subsection{Projections onto the computational grid}

Deposition patterns, as well as several parameters used in the PSPMs, are not resolved at the same scale as the computational grid. Therefore, a projection step is necessary prior to being able to integrate these quantities into the model. 

\subsubsection{Deposition patterns}

Deposition data are given for each airway generation $g_1,...,g_K$ and for a fixed set of reference particle sizes $S_1,...,S_L$, resulting in a discrete deposition pattern $(D_{k,l})$. The dose should be conserved, equivalent to conservation of number of molecules, but not number of particles.

\begin{figure}[ht!]
\begin{center}
\includegraphics[width=.7\textwidth]{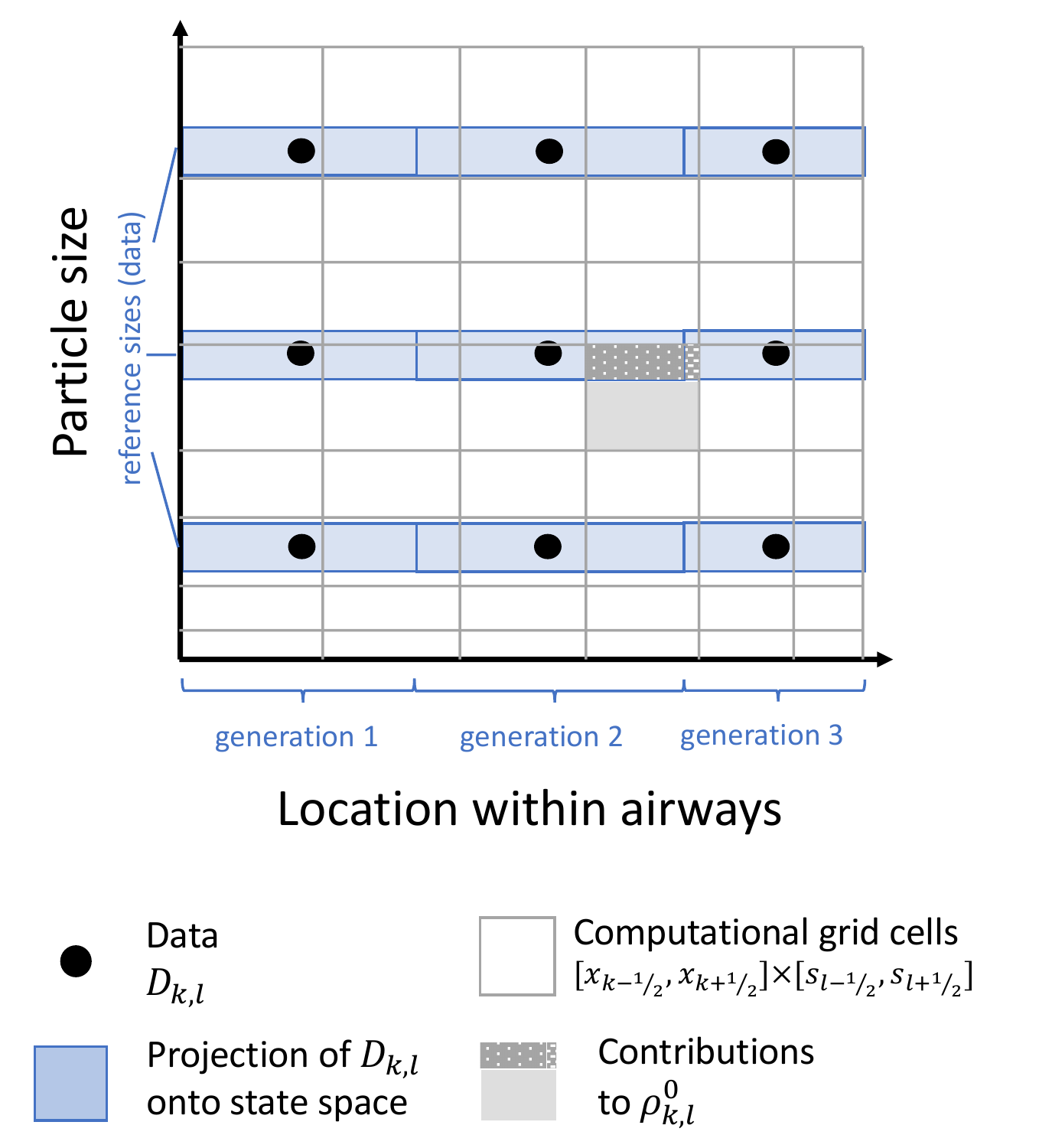}
\end{center}
\caption{ \label{fig:projection} Resolution of data against computational grid. Deposited amounts of particles with a particular size and at a particular airway location (black dots) are first distributed evenly within the respective airway generation and a small size range (blue rectangles), yielding a continuous representation of deposition within state space. The numerical approximation to the location- and size-structured density is defined on an independent computational grid. Its initial value within a grid cell is the average of the values of the continuous representation. Contributing  location-size regions to a particular grid cell are highlighted in gray.}
\end{figure}

We proceeded as follows (see Fig.~\ref{fig:projection} for an illustration):
\begin{itemize}
\item We define a region $S_k^\eps$ around $S_k$, given by$S_k^\eps=[S_k-\eps,S_k+\eps]$, with small $\eps$ such that all such regions are disjoint.
\item From the discrete values $D_{k,l}$, we define a continuous function 
\[D (x,s) = \sum_{k,l} \frac{1}{2\eps |g_k|}\mathbbm{1}_{\{x\in g_k, s\in[S_k-\eps,S_k+\eps]\}},\]
such that $\int_{g_k \times S_k^\eps} D (x,s) \text{d}x\text{d}s = D_{k,l}$.
\item We define the initial condition on the computational grid by
\[\rho^0_{k,l} = \frac{1}{\Delta x_k
\, \Delta s_l}\int\limits_{C(k,l)} D(x,s) \text{d}x\text{d}s\]
for grid cell $C(k,l) = \big[x_{k-\frac12},x_{k+\frac12}\big]\times \big[s_{l-\frac12},s_{l+\frac12}\big]$
\end{itemize}

\subsubsection{Per-generation parameters}

For a per-generation parameter (e.g., airway radius, blood flow, ...), generically denoted $P$, we construct a location-resolved representation using the previous construction only in the location coordinate, i.e.:
\begin{itemize}
\item From the discrete values $P_{k}$, we define a continuous function 
\[P(x) = \sum_{k} \frac{1}{|g_k|}\mathbbm{1}_{\{x\in g_k\}},\]
such that $\int_{g_k} P(x) \text{d}x = P_k$.
\item We define the location-resolved representation on the computational grid by
\[p_{k} = \frac{1}{\Delta x_k
}\int\limits_{x_{k-\frac12}}^{x_{k+\frac12}} P(x) \text{d}x.\]
\end{itemize}

\section{Additional model evaluations}

\subsection{Evaluation of dissolution model against in vitro data}
\label{rohrschneider}

We evaluated the dissolution model against \emph{in vitro} data from a dissolution study \cite{rohrschneider2015}, where the authors evaluated the dissolution kinetics of fluticasone propionate and budesonide particles with defined particle sizes (see Table~\ref{tab:aci}).

\begin{table}[ht!]
\begin{center}
\begin{tabular}{c|cccc}
\hline
Substance & ACI~stage & Cutoff size range & Aerodyn.~diam. & Geometric diam.\\
\hline
Fluticasone propionate & 4 & 2.1 -- 3.3 $\um$ & 2.7 $\um$ & 3.2 $\um$\\
Fluticasone propionate & 2 & 4.7 -- 5.8 $\um$ & 5.25 $\um$ & 6.2 $\um$\\
Budesonide & 4 & 2.1 -- 3.3 $\um$ & 2.7 $\um$ & 2.4 $\um$\\
\hline
\end{tabular}
\end{center}
\caption{\label{tab:aci}
Aerodynamic and geometric particle sizes corresponding to the experimental protocols of \cite{rohrschneider2015}. Particles within defined ranges of aerodynamic particle sizes were obtained from different stages of Anderson cascade impactors (ACI). For simulation of dissolution kinetics, we took the geometric diameter corresponding to the mean aerodynamic particle diameter within each impactor stage.
}
\end{table}

\begin{figure}[ht!]
\begin{center}
\includegraphics[width=\textwidth, trim = 30mm 55mm 10mm 45mm, clip]{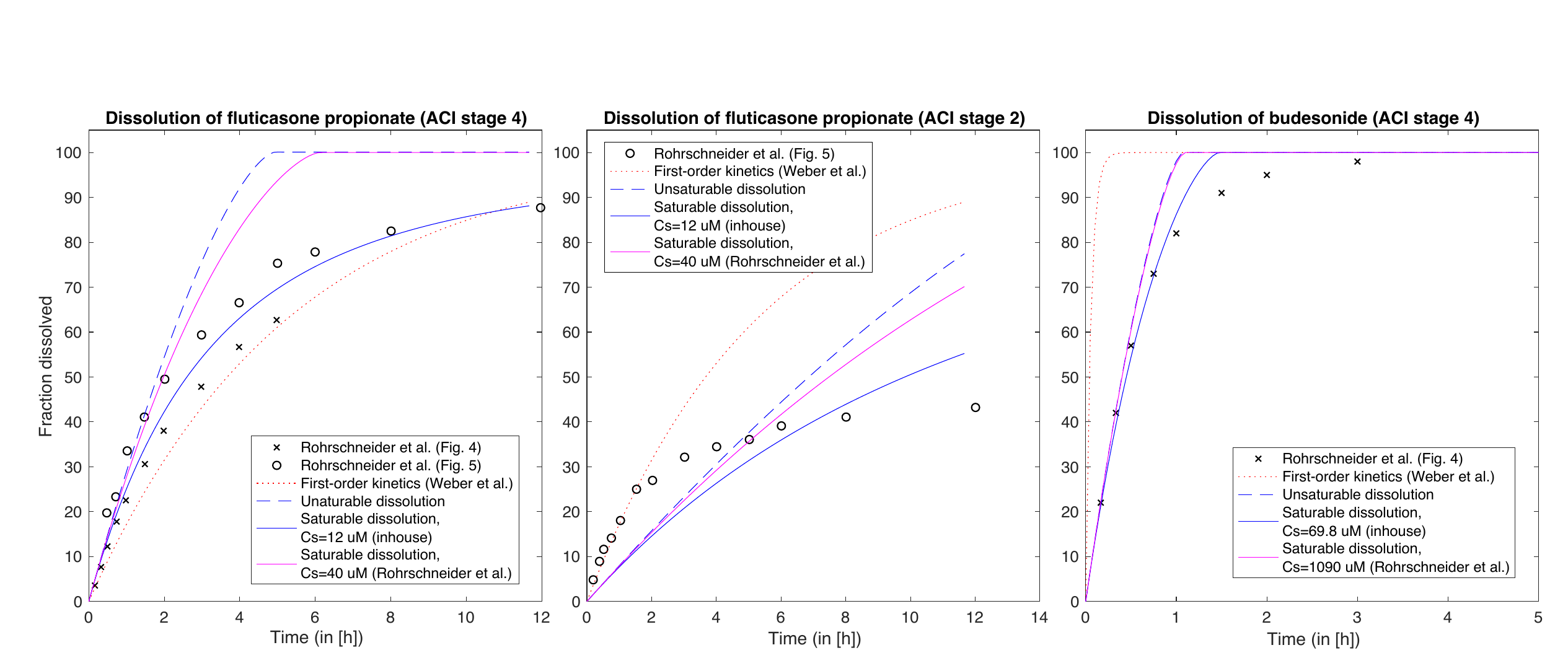}
\end{center}
\caption{ \label{fig:dissolution} Comparison of dissolution models based on in vitro dissolution data.}
\end{figure}

Based on the in vitro data, we compared different dissolution models: 
\begin{itemize}
\item a first-order dissolution model (estimated empirically; size-independent)
\item an unsaturable dissolution model (formally corresponding to $C_s=+\infty$ in the dissolution model)
\item saturable dissolution models with different solubilities 
\end{itemize}
The results are shown in Fig.~\ref{fig:dissolution}. A particle size-dependency is clearly visible, as well as a saturation effect. Among the different saturable dissolution models, the parametrization using in house data resulted in a qualitatively better description than the values reported in \cite{rohrschneider2015}.

\subsection{Representation of lining fluid height in conducting airways based on literature data}
\label{liningfluid}

\begin{figure}[ht!]
\begin{center}
\includegraphics[trim = 45 200 50 200, clip,width=.8\textwidth]{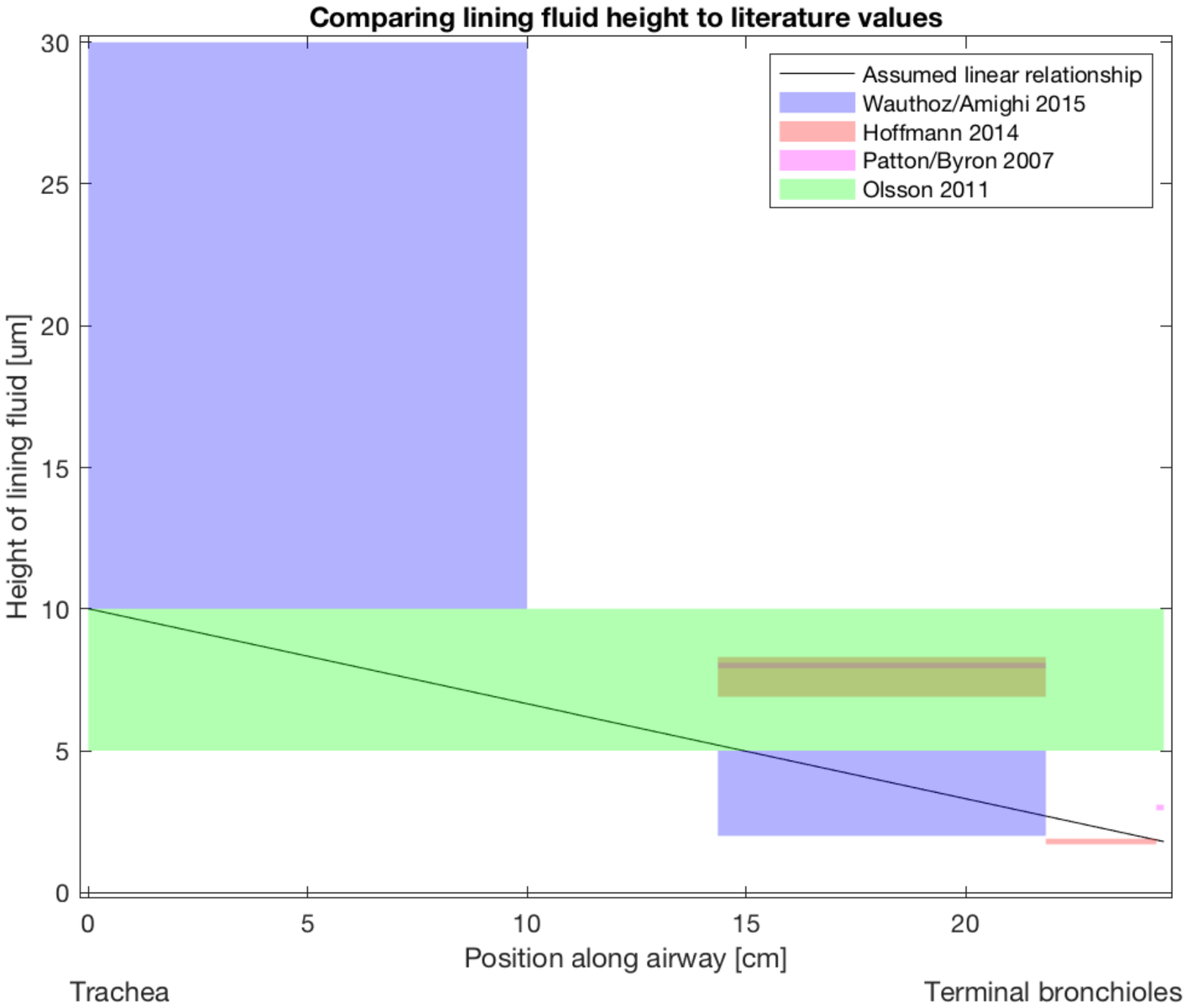}
\end{center}
\caption{ \label{fig:elf} Model for height of location-resolved lining fluid (solid black line) compared to reported literature data \cite{olsson2011,wauthoz2015,hoffmann2014,patton2007}.
}
\end{figure}

Different values for the thickness of the lining fluid layer in the conducting airways have been reported. After reviewing the literature, we concluded that the linear relationship shown in Fig.~\ref{fig:elf} adequately described the current state of knowledge. 

We decided not use literature values on total lung lining fluid volume since the reported values are not experimentally measured values but rather estimates based on height measurements and geometrical considerations. However, we note that the total lining fluid volume computed under the our geometrical assumptions ($\approx 1.2~\text{mL}$) was smaller than the ones given in the literature (10--70~mL).

\subsection{Evaluation of Usmani data}
\label{usmani}

As stated in the main text, we could not reproduce the fluticasone propionate exposure indices reported by Usmani et al.~\cite{usmani2014} based on the provided study information. Here we provide full details for this statement. For the smallest particles of 1.5 $\um$ diameter, Usmani et al.~reported an $\text{AUC}_\text{0-12h}$ of $923.28~\text{pg}\cdot\text{h}/\text{mL}$, i.e. in molar units $1.84~\text{nM}\cdot\text{h}$. Assuming 100 \% lung uptake, no mucociliary clearance and a full systemic uptake within 12~h, and taking the literature value for fluticasone propionate clearance of $73~\text{L}/\text{h}$ \cite{weber2013}, we obtain a very conservative upper bound of
$\text{AUC}_\text{max} = \frac{\text{Dose}}{\text{CL}\cdot \text{MW}} = 1.37~\text{nM}\cdot\text{h}$. 

A more realistic, albeit still conservative calculation and a simulation with the PDE model are shown in Table~\ref{tab:usmani}. In conclusion, the reported AUC value is approximately 2-4 times larger than what could be reasonably expected. Accordingly, $C_\text{max}$ values are also much higher than predicted by the PDE model.

\begin{table}[htp]
\begin{center}
\begin{tabular}{ccc|c}
 \hline
\multicolumn{3}{c|}{Assumptions for AUC calculation} & Calculated AUC\\
\cline{1-3}
Lung dose & MCC & Timespan & (compared to reported AUC)\\
 \hline
 100\% & no  & $\text{AUC}_\text{0-$\infty$}$ & 26\% lower than reported\\
 56\%  & no  & $\text{AUC}_\text{0-$\infty$}$ & 58\% lower than reported\\
 56\%  & yes & $\text{AUC}_\text{0-12h}$  &  74\% lower than reported\\
\end{tabular}
\end{center}
\caption{\label{tab:usmani}
Comparison of calculated AUC vs.~reported AUC$_\text{0-12h}$ for different assumptions. Even under the most conservative assumptions, AUC is considerably underestimated, which becomes more pronounced as the model gets more realistic.
}
\end{table}

\section{Simulation of pulmonary deposition patterns}

\subsection{Settings in MPPD software}
\label{mppd-healthy}

In order to predict the pulmonary deposition patterns, the MPPD software v2.1 was used \cite{mppd}. This software allows to predict the generation-dependent pulmonary deposition of inhaled particles, where generations~1-17 represent the conducting airways (generation 1 = trachea) and generations~18-25 the alveolar space. 
Three types of input data are required in the MPPD software: (1) airway morphometry, (2) particle properties, and (3) exposure condition, as outlined below. The MPPD software was only applied to simulate the deposition patterns but not used to investigate the clearance of particles from the lung.

\begin{description}
\item[Airway morphometry.]
For all predictions performed with the MPPD software, the airway morphometry was represented by the human ``Yeh/Schum 5-Lobe'' model \cite{yeh1980}. The inhalation flow characteristics were assumed to be represented by uniform expansion of the lung so that consequently also the inhalation and exhalation flow were constant over time. The standard airway morphometry defined in the MPPD software was selected for all deposition pattern predictions, i.e. the default values for functional residual capacity (3300~mL) and upper respiratory tract volume (50~mL) were used \cite{icrp1994}. 

\item[Particle properties.]
The inhaled particle properties were defined based on the information in the respective publications, or alternatively for the respective inhalation device (references are provided in Table~\ref{tab:mppd}). For all predictions, the density of the particles was set to $1~\text{g}/\text{cm}^3$; and the particle diameter was defined as the mass median aerodynamic diameter, which is typically provided in literature. As described in the main manuscript, the difference between  aerodynamic and geometric diameters was accounted for, such that the real surface area could be used as an input parameter to the dissolution model. The MPPD software was only used to predict  pulmonary deposition patterns of monodisperse particles. To predict the deposition patterns for the monodisperse gold/polystyrene particles (Study~I) and the inhaled monodisperse fluticasone propionate particles (Study~II), this information was sufficient. Whenever pulmonary deposition patterns of a particle size distribution were required (Studies~III/IV), these were generated in a two-step approach. First, all relevant monodisperse particles size bins of the particle size distribution were simulated as monodisperse particles with the MPPD software. In a second step, the complete deposition pattern was calculated by normalizing the deposited amount per particle size bin by the dose in this respective bin. The two additional options of the MPPD software, namely the ``Nanoparticle Model'' and ``Inhalability Adjustment'' were not applied to predict the deposition patterns.

\item[Exposure conditions.]
The exposure scenario was set to constant exposure and the body orientation during the inhalation process was assumed ``upright''. Furthermore, for all predictions, it was assumed that the breathing scenario was represented by oral breathing, which is the typical inhalation route for drugs delivered to the lungs. Breathing frequency, tidal volume, inspiratory fraction as well as pause fraction were all defined based on the inhalation flow properties provided in the respective publications (see Table~\ref{tab:mppd}).
\end{description}

\begin{table}[htp]
\begin{center}
\begin{tabular}{R{3.4cm}C{2.5cm}C{2.5cm}C{2.5cm}C{2.5cm}}
 \hline
& {\bf Study} I & {\bf Study II} & {\bf Study III} & {\bf Study IV} \\ 
\hline
{\bf Particle properties} & & & &\\
Substance & gold / polystyrene & fluticasone propionate & fluticasone propionate & budesonide\\
Formulation type & monodisperse & monodisperse & polydisperse & polydisperse \\
Particle size(s)  & $5 \um$ diameter & 1.5 / 3  / 6 $\um$ diameter & distribution based on \cite{tamura2012} & distribution based on \cite{tamura2012}\\  
\hline
{\bf Exposure scenario} & & & &\\
Device & custom setup (see \cite{smith2008}) & Inhalation chamber & \Diskus & \Turbohaler\\ 
Breathing frequency & 6/min & 5/min & 6/min &6/min \\
Tidal volume & 200~mL & 2000~mL & 2000~mL & 2000~mL \\
Inhalation time & 1~sec & 4~sec  & 1.33~sec & 1.33~sec\\
Exhalation time & 1~sec & 3~sec  & 2.67~sec & 2.67~sec\\
Pause time & 8~sec & 5~sec  & 6~sec & 6~sec\\
Inhalation flow & 12~L/min & 30~L/min & 90~L/min & 60~L/min\\
\hline
{\bf Deposition pattern corrections} & & & &\\
Lung dose & no correction & 56.3\% / 51\% / 46.0\% & 14.5\% based on \cite{backman2016} & 35\% based on \cite{backman2016}\\
Central/peripheral deposition ratio & no correction & central deposited fraction: 
56.1\% / 65.7\% / 75.4\% & 2-fold lower alveolar deposition for asthma patients \cite{usmani2005} & 2-fold lower alveolar deposition for asthma patients \cite{usmani2005}\\
\hline
\end{tabular}
\end{center}
\caption{\label{tab:mppd}
Study-specific input data to the MPPD software.
}
\end{table}

\subsection{Adaptation of deposition patterns for asthmatic patients}
\label{mppd-asthma}

Since the MPPD software predicts deposition patterns in healthy volunteers, it cannot directly be used to predict deposition patterns in asthmatic or COPD patients. In these patients, due to narrowed airways, deposition is more central in comparison to healthy volunteers. Whenever patients were considered in a study rather than healthy volunteers, deposition patterns had to be adapted adequately. To this end, the fraction of the inhaled dose deposited in any specific airway generation was increased by an adjustment factor such that the deposited fraction of the lung dose in the alveolar space was 2-fold lower than in healthy volunteers. This number was derived from published data on conducting airway to alveolar deposition ratios \cite{usmani2005}.

\section{Generation of in-house data}

\subsection{In vitro solubility determination in surfactant containing medium}

For \emph{in vivo} relevant characterization of the drug solubility, the surfactant-containing medium \Alveofact, a commercially available product, was taken. \Alveofact contains phospholipids obtained from bovine lung (i.e., surfactants) and is available as dry powder ampoules ready for reconstitution. 
As reconstitution medium, a 0.1 mol/l sodium dihydrogenecarboante buffer with pH 7.4 was used. A suspension with 50 mg/ml \Alveofact was produced according to the information and instruction for use of the commercial product. At these concentrations, \Alveofact forms a micellar system.
1 mg of drug (either budesonide or fluticasone propionate) is suspended in 1 ml of this medium and shaken for 24 h at 37 $^\circ$C. Afterwards, the suspension is filtered with a commercially available Whatman Mini-UniPrep syringeless filter containing a 0.45 $\mu$m filter membrane out of glass microfibers. As the micelles pass this membrane and as the concentration of phospholipids is too high to be directly injected in the HPLC system for analysis of the solubilized amount of drug, the micelles are destroyed by adding DMSO in a 1:1 ratio to the filtered micellar solution. The phospholipids can be separated by an additional 5 -- 10 minutes centrifugation step. 
A small aliquot of the remaining solution is taken and injected into a HPLC system for quantitative analysis of the solubilized amount of drug.

\subsection{Blood to plasma ratio determination}

To determine the Blood:Plasma (BP) ratio, the respective amount of the drug (i.e., fluticasone propionate) was added to 490 $\mu$L human blood and to 490 $\mu$L plasma samples to obtain a drug concentration of 10 $\mu$M. Both the plasma (plasma sample \#2) and the blood samples were incubated with the drug for 15 minutes at 37 $^\circ$C (n=3). 
Afterwards the blood sample was centrifuged at 3000 rpm to separate the blood cells from the plasma sample (plasma sample \#1). The respective plasma concentrations were determined by MS-based analysis.
In a last step, the BP ratio was calculated by dividing the drug concentration in plasma sample \#2 by the drug concentration in plasma sample \#1. To ensure quality of the measurement, the degree of hemolysis was determined and considered negligible for all BP experiments. In addition, a control experiment without any drug was performed in parallel to determine the hematocrit of all three samples.

